\documentclass{aa}  
\usepackage{graphicx}
\usepackage{txfonts}
\newcommand \htwo  {\mbox{H$_2$}}

\newcommand \degree {\mbox{$^\circ$}}
\newcommand \msun {\mbox{$\mathcal{M}_{\odot}$}}

\newcommand \kms {\mbox{km~s$^{-1}$}}
\newcommand \kmskpc {\mbox{km~s$^{-1}$~kpc$^{-1}$}}
\newcommand \micron  {\mbox{$\mu$m}}
\newcommand \jyperbeam {\mbox{Jy~beam$^{-1}$}}
\newcommand \mjyperbeam {\mbox{mJy~beam$^{-1}$}}
\newcommand \twelvecoonezero {\mbox{$^{12}$CO(1--0) }}
\newcommand \twelvecotwoone  {\mbox{$^{12}$CO(2--1) }}
\newcommand \twelvecothreetwo  {\mbox{$^{12}$CO(3--2) }}
\newcommand \angstrom {$\AA$}
\newcommand \HS {{\bf{HS06}}}

\begin{document}

   \title{Nuclear starburst-driven evolution of the central region \\in NGC 6764}


   \author{S. Leon,
          \inst{1}
           A. Eckart, \inst{2}
	   S. Laine, \inst{3}
	   J. K. Kotilainen, \inst{4}
           E. Schinnerer, \inst{5}
           S.--W. Lee, \inst{6}
	   M. Krips, \inst{2}
           J. Reunanen, \inst{7} \\
	   and J. Scharw\"{a}chter \inst{8}
}

      \offprints{S. Leon, leon@iram.es}
      
\institute{$^1$ Instituto de Radio Astronom\'{\i}a Milim\'{e}trica, Granada, Spain \\
	      $^2$ University of Cologne, I. Physikalisches Institut, Cologne, Germany\\
	      $^3$ Spitzer Science Center, Caltech, Pasadena, USA\\
	      $^4$ Tuorla Observatory, University of Turku,  Piikki\"o, Finland\\
              $^5$ Max-Planck-Institut f\"{u}r Astronomie, Heidelberg, Germany\\
	      $^6$ University of Toronto, Astronomy Department, Toronto, Canada\\
	      $^7$ University of Leiden, Department of Astronomy, Leiden, The
	      Netherlands \\
	      $^8$ European Southern Observatory, Santiago, Chile
              }

   \date{Received XX; accepted XX}

 
  \abstract
   {}
   {We study the CO and the radiocontinuum emission in an active galaxy to analyze the interplay
between the central activity and the molecular gas.}
   {We present new high-resolution observations of the \twelvecoonezero
and \twelvecotwoone emission lines, and 3.5 cm and 20 cm radio continuum emission 
in the central region of the LINER/starburst galaxy NGC~6764.}
   {The galaxy has
an outflow morphology in radio continuum, spatially coincident with 
the CO and H$\alpha$ emission, and centered slightly off the 
radio continuum peak at the LINER nucleus. The total molecular gas mass in the 
center is about 7$\times10^{8}$ \msun, using a CO luminosity to total molecular
gas conversion factor that is three times lower than the standard one. 
\twelvecoonezero emission is found near the boundaries of the radio continuum 
emission cone. The outflow has a projected expansion velocity of 25 \kms 
relative to the systemic velocity of NGC~6764. About $4\times 10^{6}$ \msun\ of 
molecular gas is detected in the outflow.  The approximate location 
($\sim$ 1 kpc) of the dynamical inner Lindblad resonance has been derived from 
the rotation curve. The peak of the CO emission is slightly ($< 200$ pc) 
offset from the peak of the radio continuum.}
   { The molecular gas has most likely been ejected 
by the stellar winds from the recent starburst, but the CO line ratios show indication of an interaction with the AGN.
The energy released by the nuclear starburst is sufficient to explain the observed outflow, even if the data cannot exclude the
AGN from being the major energy source. Comparison of the outflow 
with hydrodynamical simulations suggests that the nuclear starburst is 3--7 Myr 
old and the bubble-like outflow is still confined and not freely expanding.}


	\titlerunning{Molecular gas in NGC~6764}
 	\authorrunning{Leon et al.}
   
\maketitle
%

\section{Introduction}
The nuclear region surrounding active galactic nuclei (AGN) often has 
the highest level of molecular gas concentration and star formation in galaxies. 
The processes that drive the excitation and kinematics of the circumnuclear gas, 
namely, the abundant star formation and the jets and photoionisation associated 
with the AGN, influence the circumnuclear gas in several ways. 
The onset and 
shutdown of the starburst are driven by the properties of the molecular gas, 
including its
density and velocity dispersion, by gravitational instabilities (e.g., 
Combes 2001), and by processes that tend to prevent star formation, such as the 
local tidal field (Launhardt, Zylka \& Mezger 2002), starburst feedback (e.g., 
Dopita 1985, Scalo \& Chapell 1999), gas consumption (e.g., Franceschini et al. 
1998), and tidal shear (e.g., Kenney et al. 1993).
Fueling of starbursts is facilitated by nonaxisymmetric components such as bars, and by tidal 
torques from interactions and mergers (e.g. Combes \& Gerin 1985, Athanassoula 1992, 
Mihos \& Hernquist 1994, Laine et al. 2002)

The formation of powerful gas and plasma outflows from the center of disk 
galaxies provides the basis for the metal enrichment of galaxy halos (e.g.,
Kunth et al. 2002). Several mechanisms have been proposed to explain these 
outflows. The first possibility is a thermal wind from a circumnuclear 
starburst (e.g., Colina, Arribas \& Borne 1999). The second possibility is 
a thermal wind from an AGN, as suggested by, e.g., Krolik \& Bergelman (1986).
Finally, it is possible that ram-pressure along the radio jet is driving the
outflow (e.g., Taylor, Dyson \& Axon 1992). Large-scale radio halos outside 
the galaxy disk, which most likely result from plasma and magnetic fields that 
have been blown off the disk by one of the above-mentioned mechanisms, have 
been known for more than two decades (e.g., Hummel, van Gorkom \& Kotanyi 1983, 
Hummel, Beck \& Dettmar 1991). 

The nuclear starburst leads to the formation of OB associations or super star
clusters, from which the massive stars, through strong supersonic stellar 
winds, inject both energy and mass into the surrounding interstellar medium. 
The formation of an expanding starburst ``superbubble'' will be accompanied by 
extended X-ray emission from hot gas, and optical H$\alpha$ line emission 
from the cooler shell. The superbubble that has a larger internal ISM pressure
than its surroundings expands into the low-pressure halo ISM until the onset of 
Rayleigh-Taylor instabilities, which disrupt the superbubble and allow the hot 
gas to escape in the form of a galactic wind, as is seen, e.g., in M82 
(Strickland, Ponman \& Stevens 1997). The cold molecular gas is expected to 
be compressed around the superbubble into an expanding cocoon.

Molecular gas in outflows has been detected in a few galaxies. 
M82 has an outflow bubble expanding at 45 \kms (Wei\ss\ et al. 1999), 
with a molecular outflow gas mass of $\sim 10^6$ \msun .
NGC~2782 harbors an outflow with
about $\sim 2\times 10^7$ \msun\ of molecular gas (Jogee et al. 1998). The
outflow in NGC~2782 appears to be in the early stages (its age is $\sim 4\times 
10^6$ years), so that it is seen before the onset of Rayleigh-Taylor 
instabilities. In NGC~3079 four expanding CO shells have been identified 
(Irwin \& Sofue 1996). These appear to have originated from a single starburst 
that took place a few Myrs ago. 

The behavior of the molecular gas in such 
energetic outflows is still poorly understood. The kinematic and potential 
energies of the molecular gas are part of the total energy budget of the outflow, 
and one
can use the observed properties of the gas (its location, mass and velocity) to
constrain the origin of the outflow.
Recently Sakamoto et al. (2006) observed molecular gas shells with the Submillimeter Array (SMA) 
in the circumnuclear disk of NGC~253.
Further constraints on the nature and
location of the energy source can be provided by the current gas phase, as in 
NGC~2782 (Jogee et al. 1998), or an unusual geometry of the outflow, such as
the off-center outflow in NGC~5775 (Lee et al. 2001). Since gas is fuel for 
star formation, when it falls back onto the galaxy disk it may induce 
new episodes of star formation (e.g., Norman \& Ikeuchi 1989).

NGC~6764 is a barred spiral with an optically classified 
low-ionisation nuclear emission-line region (LINER) nucleus 
(Osterbrock \& Cohen 1982). 
This nearby active galaxy (D=32 Mpc; 1\arcsec~=~160 pc) 
hosts a very young nuclear starburst 
as shown by its Wolf--Rayet (W-R) feature
at 0.47 \micron\ (HeII; Schinnerer et al. 2000). 
Eckart et al.
(1991, 1996) and Schinnerer et al. (2000) analyzed the central starburst, 
traced by the 
W--R emission, in great detail. The millimetre observations have revealed a 
central concentration  of molecular gas which also emits near-infrared line 
emission. The 2.12 \micron\ H$_2$ line has features that are also seen in
radio continuum. A population synthesis model based on 
NIR spectral lines suggests that NGC~6764 
has undergone two recent starbursts. One of them took place 3--5 Myr ago,
producing W--R features, and another 
15--50 Myr ago. {\it ROSAT} X-ray data trace mainly the central AGN 
which varies by a factor of two over seven days (Schinnerer et al. 2000). By 
studying HCN and CO emission in the center 
of NGC~6764, Kohno et al. (2001) classified this galaxy as a combined
AGN/starburst, based on the high HCN/CO line
ratio ($\sim 0.7$) in the nucleus which suggests the presence of nuclear activity. 
Recently Hota \& Saikia (2006, hereafter \HS) presented GMRT+VLA radiocontinuum  and HI observations
together with H$\alpha$ data of the galaxy NGC 6764 where they found bipolar radio bubbles along the
minor axis together with a compact core. Their HI observations reveal two peaks of emission at the ends
of the bar and a depletion in the center  together an HI absorption towards the radio core. 
The coexistence of an AGN and a starburst 
nucleus in NGC~6764 makes this galaxy an ideal target for studying the influence
of both processes.

We present new high spatial resolution 
observations of the \twelvecoonezero and \twelvecotwoone lines 
made with the IRAM Plateau de Bure Interferometer (PdBI\footnote{Based on
observations carried out with the IRAM Plateau de Bure Interferometer. IRAM is 
operated by CNRS (France), the MPG (Germany) and the IGN (Spain).
}).
We also present high resolution 
VLA\footnote{The VLA is a research facility of the National Radio Astronomy 
Observatory (NRAO), which is operated by Associated Universities Inc., under 
contract with the National Science Foundation.} radio continuum maps at 3.5 cm 
and 20 cm (Laine et al. 2006). 
We describe our observations in \S 2. In \S 3 we discuss the 
thermal and synchrotron emission associated with the hot ISM, and their
connection to the nuclear star formation. The distribution, kinematics, 
physical properties, and mass of the molecular gas in the outflow in NGC~6764
are described in \S 4. The energies involved in the outflow as determined by the
molecular gas and radio continuum observations of the outflow are addressed
in \S 5. We discuss our results in \S 6 and summarize them in \S 7.
Throughout this paper, we use the following cosmological parameters: $H_0 = 75~\kms$, 
$\Omega_\Lambda = 0.7$ and $\Omega_b = 0.3$

\begin{table}
\caption{Global properties of NGC~6764.
}
\label{tab_glob_param}
\begin{tabular}{lll}
\hline
Property & Value  & References\\
\hline
\hline
Hubble type  & SB(s)bc  &  NED \\
Distance (Mpc)  &  32 &  H$_0=75$ \kms~Mpc$^{-1}$ \\
Inclination  & 61.5\degree  & LEDA \\
Major Axis P.A.   & 62\degree  & LEDA\\
Bar P.A.  & 74.6\degree & LEDA \\
S$_{12\micron}$ (Jy) &  0.36  & NED\\
S$_{25\micron}$ (Jy) &  1.29  & NED\\
S$_{60\micron}$ (Jy) &  6.33  & NED\\
S$_{100\micron}$ (Jy) & 11.56 & NED\\
L$_{\mbox{FIR}}$ (10$^{10} $ L$_\odot$) & 1.1 &   \\
L$_{\mbox{H$\alpha$}}$ (10$^{40}$ erg~s$^{-1}$)  & 23 &  \\
M$_{\mbox{HI}}$ (10$^9$\msun)  & 7.0 & NED\\
S$_{90cm}$ (mJy) & 309 & WENSS\\
S$_{80cm}$ (mJy) & 244 & Douglas et al. (1996)\\
S$_{20cm}$ (mJy) & 133 & White \& Becker (1992) \\
S$_{6cm}$ (mJy) & 34 & Becker, White \& Edwards (1991) \\ 
S$_{3.5cm}$ (mJy) & 31 & This paper \\
\hline
\end{tabular}
\end{table}

\section{Observations and data reduction}

\subsection{Radio continuum data}

\begin{figure*}
\resizebox{17cm}{8.5cm}{\includegraphics{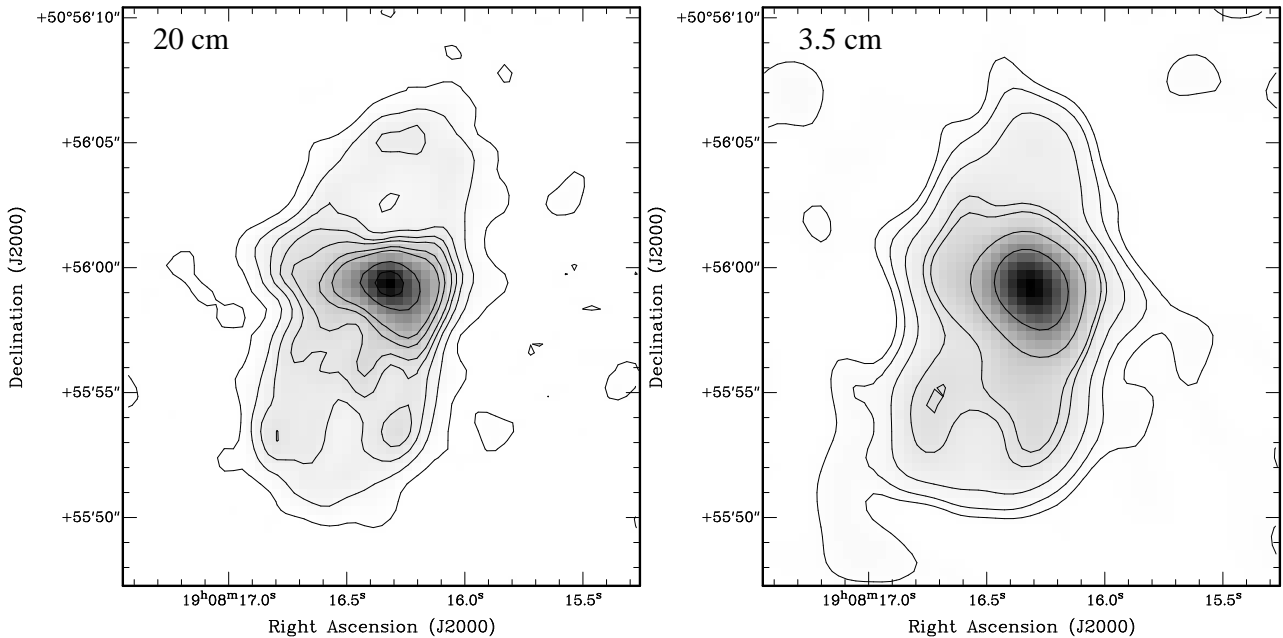}}
\caption{Left: radio continuum emission at 20 cm with a synthesized beam of 
$1\farcs 39\times 1\farcs 26$. The contour levels are at (0.1, 0.5, 1, 1.5, 2, 3, 4, 8, 12) mJy~beam$^{-1}$ and the rms noise is 70 $\mu$Jy~beam$^{-1}$. Right: radio continuum emission at 3.5 cm (VLA) with a synthesized beam of 
$2\farcs 50\times 2\farcs 0$. The contour levels are at (0.1, 0.2, 0.4, 0.8, 1, 2, 4, 8) mJy~beam$^{-1}$ and the rms noise is 56 $\mu$Jy~beam$^{-1}$.}
\label{fig_vla20_35}
\end{figure*}

The radio continuum data were obtained with the Very Large Array (VLA) at 
3.5 and 20 cm  in C and A configurations (see Figure~\ref{fig_vla20_35}),
respectively. We obtained a comparable spatial resolution at the two observed
wavelengths, $1\farcs 39\times 1\farcs 26$ (P.A.=$-45$\degree ) at 20 cm and 
$2\farcs 50\times 2\arcsec$ (P.A.=15.5\degree) at 3.5 cm. For further details
about the radio continuum observations, see Laine et al. (2006).
The peak position of the 20 cm emission, $\alpha_{20cm}(J2000)$~=~19:08:16.316, 
$\delta_{20cm}(J2000)$~=~50:55:59.37, was determined to an accuracy 
better than 0\farcs 1, by filtering out 
large spatial frequencies, using a wavelet transformation (e.g., Leon et al. 2000), 
and fitting a Gaussian to the remaining emission. 

\subsection{CO data}
The \twelvecoonezero and \twelvecotwoone lines were observed with the 
IRAM interferometer (see Table \ref{tab_obs_pdbi}), located on the Plateau de Bure (France), in the 
B and C configurations of the array, in 1998. We calibrated the visibilities using the
radio galaxy  3C~380 as an amplitude calibrator and 1823+568 and 3C~380 as 
phase references. The Plateau de Bure Interferometer (PdBI)  was phase-centered
at $\alpha_{J2000}$=19:08:16.370 and $\delta_{J2000}$=+50:55:59.58. 
The velocity resolution was set at 10 \kms\ for the two 
CO transitions, and the CLEANed maps have been restored with a synthesized beam 
of $2\farcs 0\times 1\farcs 6$ [P.A.=7\degree; \twelvecoonezero],
and $1\farcs 25\times 1\farcs 25$ [P.A.=0\degree; \twelvecotwoone]. The rms 
noise for the channel maps is 2 mJy~beam$^{-1}$ [58 mK; 
\twelvecoonezero], and 5.5 mJy~beam$^{-1}$ [82.5 mK; 
\twelvecotwoone]. Hereafter the velocity reference is fixed to 2400 \kms, 
giving a redshifted frequency of 114.3489 GHz for the \twelvecoonezero 
observations, and 228.6932 GHz for the \twelvecotwoone observations.

The total integrated \twelvecoonezero flux is 225 Jy~\kms in the 55\arcsec\ beam of 
the NRAO 12-m telescope (Sanders \& Mirabel 1985). 
Our PdBI \twelvecoonezero observations detected a total flux of 112 Jy~\kms, i.e., 
filtering out about 50\% of the total flux, assuming all the emission is 
concentrated in the 45\arcsec\ primary beam of the PdBI. This implies that a 
large fraction of the molecular gas must be in an extended smooth component 
to which the interferometer is insensitive. From Eckart et al. (1991) we estimate with IRAM-30m observations the 
\twelvecotwoone  integrated intensity to be $\sim 480$ Jy~\kms in the center of NGC~6764, which implies
 a filtering of about 65\% of the \twelvecotwoone flux  by the PdBI observations.

\begin{figure*}
\begin{center}
\resizebox{17cm}{!}{\includegraphics{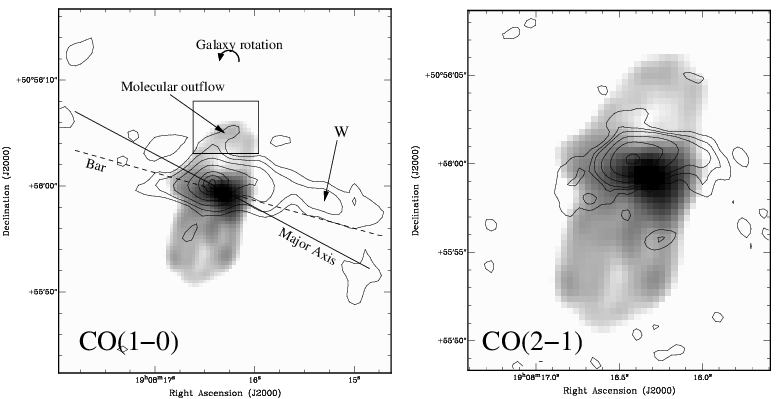}}
\caption{Left: VLA 20 cm radio continuum emission (greyscale) overlaid on the 
integrated intensity image of the \twelvecoonezero emission. 
Right: \twelvecotwoone integrated intensity image. The contours for 
\twelvecoonezero are at (1, 2, 4, 8, 15, 20, 25) \jyperbeam~\kms, and for \twelvecotwoone at (2, 4, 6, 12, 24, 52,) 
\jyperbeam~\kms. The position of the major axis of the galaxy and of the bar 
are shown by solid and dashed lines, respectively.
}\label{fig_ico_pdbi}
\end{center}
\end{figure*}

The integrated intensity maps for 
the \twelvecoonezero\ and \twelvecotwoone\ emission lines are shown in 
Figure~\ref{fig_ico_pdbi}. The position of the CO emission peak 
$\alpha_{CO}(J2000)~=~$19:08:16.402, $\delta_{CO}(J2000)~=~$50:56:0.29 has been 
estimated using the higher spatial resolution of the \twelvecotwoone transition.
Assuming that the spiral arms are trailing, and noting that the receding 
velocities in the disk are on the eastern side of NGC~6764, the northern side 
of the galaxy disk is closer to us.

\begin{table}
\caption{PdBI observations
}
\begin{tabular}{l|ll}
Parameter  &  \twelvecoonezero & \twelvecotwoone \\
\hline
\hline
Field of View    &  45\arcsec (6.9 kpc)   & 22\arcsec  (3.4 kpc)\\
Bandwidth	       &  270 MHz  		& 473 MHz   \\
Velocity resolution	& 10 \kms		& 10 \kms  \\
Synthesized beam size &	$2\arcsec \times 1\farcs 6$, P.A. 7\degree &  $1\farcs 25 \times 1\farcs 25$ \\
rms noise (per channel) & 2 \mjyperbeam  & 6 \mjyperbeam \\
\hline
\end{tabular}
\label{tab_obs_pdbi}
\end{table}

\section{Thermal and synchrotron emission}

\subsection{Radio continuum}

The total integrated 20 cm flux density from the VLA observations is 90 mJy, 
which compared to the single dish flux density of 133 mJy (White \& Becker 1992),
implies that $\sim$70\% of the total flux density is recovered by 
the VLA observations. The total 3.5 cm flux density is 31 mJy. 
Both VLA maps at 20 and 3.5 cm present the same features with a better spatial resolution 
at 20 cm (see Fig. \ref{fig_vla20_35}). A bright peak is observed towards the nucleus of NGC~6764. 
Two 5\arcsec\ spurs emerge from the unresolved nucleus, towards the south and the south-east. The emission near the nucleus is not consistent with a just a nuclear point source but hints at the presence of a nuclear radio jet, as proposed by \HS.
The most remarkable feature
is the shell-like emission, at both wavelengths, centered on the central peak with its major axis
perpendicular to the major axis of NGC~6764 and extending about 1 kpc above and below the disk plane of the highly inclined
galaxy NGC~6764. The clumpy shell emission is brighter towards south than towards north. These radio bubbles were previously reported
by Baum et al. (1993), \HS\ and Laine et al. (2006).



\subsection{H$\alpha$ emission}

\begin{figure}
\resizebox{8.5cm}{8.5cm}{\includegraphics{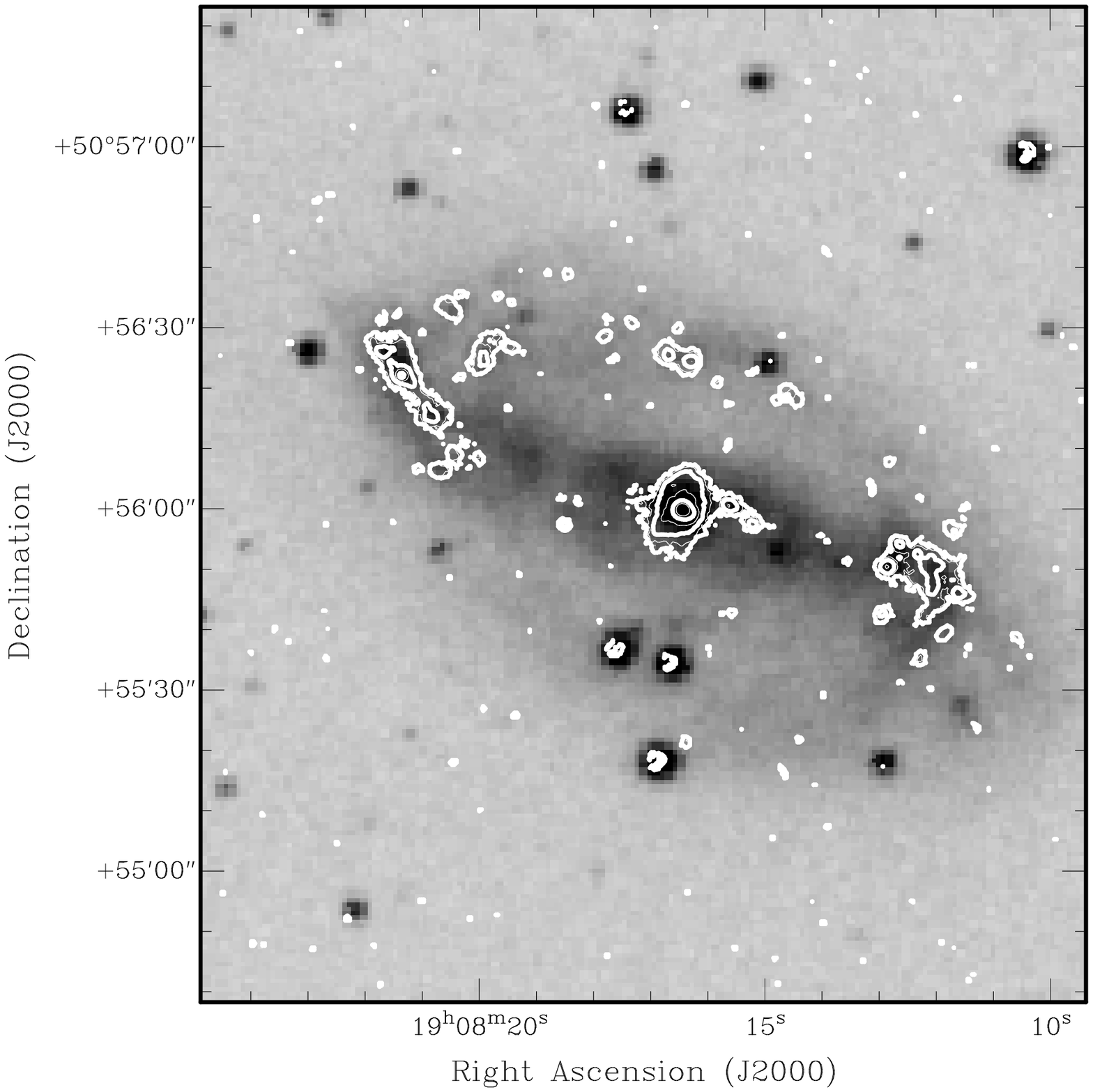}}
\caption{B image (from Digital Sky Survey) of NGC~6764 overlaid by
H$\alpha$ emission contours. The contours are 
spaced by 0.5 dex steps, starting at the 0.5 $\sigma$ level.}
\label{fig_halpha_dss}
\end{figure}
\begin{figure}
\resizebox{8.5cm}{8.5cm}{\includegraphics{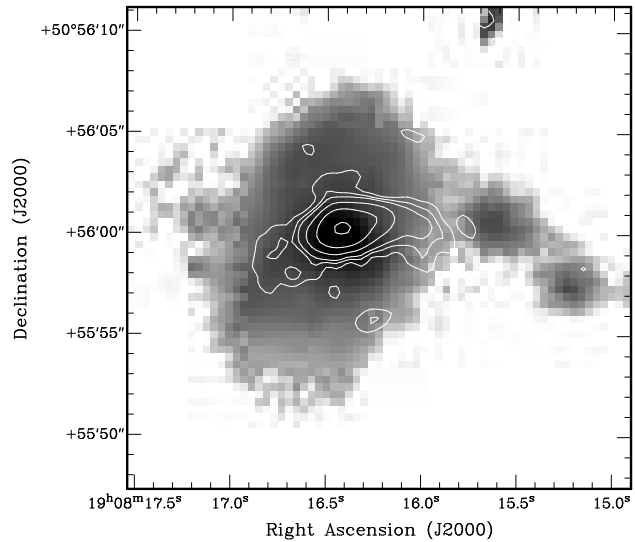}}
\caption{H$\alpha$ emission (greyscale) overlaid by the \twelvecotwoone emission contours (same levels as in 
Fig. \ref{fig_ico_pdbi}).}
\label{fig_halpha_co21}
\end{figure}

An H$\alpha$ image of NGC~6764 was kindly given to us by D. Frayer. The observations were done
at the Palomar 60-inch telescope with a seeing of 1--3\arcsec, using the X-filter centered at 6616~\angstrom
with a width of 20~\angstrom, leaving out most of the [NII] emission line.
Figure~\ref{fig_halpha_dss} shows the H$\alpha$ emission at the center of 
NGC~6764. Apart from the most intense H$\alpha$ emission in the central
region,  two other regions of emission are seen near the tip of the bar, which
also lies near the beginning of the spiral arms. 
The most striking characteristic
of the H$\alpha$ emission near the center of NGC~6764 is its bubble-like 
shape, but its size is smaller than the size of the radio continuum outflow
feature in the south. In Figure~\ref{fig_halpha_co21} a 4\arcsec\ spur is 
observed to the southeast of the location of the \twelvecotwoone peak, bending towards 
south. This spur is spatially correlated with faint ($<6~\jyperbeam~\kms$) 
\twelvecotwoone emission.

\begin{figure}
\resizebox{9cm}{8.5cm}{\includegraphics{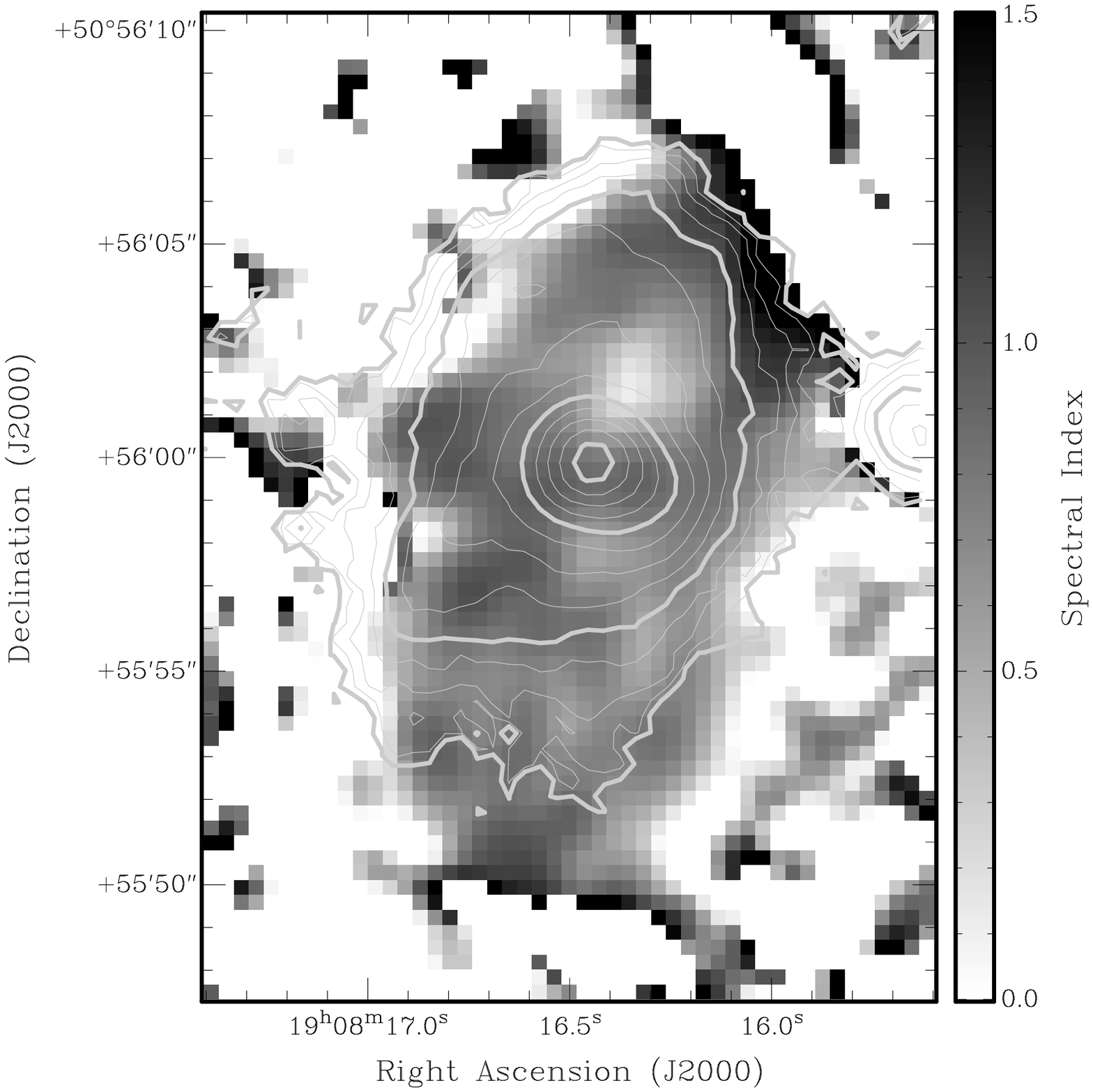}}
\caption{Spectral index $-\alpha$ between 3.5 cm and 20 cm (greyscale), overlaid by 
H$\alpha$ emission contours.}
\label{fig_halpha_vlaspectralindex}
\end{figure}

In Figure~\ref{fig_halpha_vlaspectralindex} the H$\alpha$ emission is compared 
with the radio continuum spectral index. The ``hot spot'' in the bubble cavity 
with a flat spectral index ($\alpha \sim -0.2$) is associated with H$\alpha$ 
emission driven by the nuclear starburst. Similarly, 
the tip of the southeastern H$\alpha$ emission spur coincides with a small 
``hot spot'' that has a flat spectrum.  

\subsection{Nuclear star formation}


The nuclear star formation history was studied in detail by Schinnerer et al. (2000) using near-infrared K-band imaging spectroscopy of the very center of NGC~6764. To get the current star formation rate we have used here the FIR and the H$\alpha$ emission line. We estimate as well the contribution of the thermal and non-thermal radiocontiuum emission.

\subsubsection{Star formation rate}

The total FIR luminosity is computed for the whole galaxy from the 60 and 100 \micron\ IRAS fluxes (see Table \ref{tab_glob_param}), 
but is very close to the bolometric value $L_{bol}=9\times 10^9$ $L_\odot$ computed for the nuclear region by Schinnerer et al (2000). Following Condon (1992), the star formation rate (SFR) for massive stars is
$SFR~(M>5~\msun) = 9.1\times10^{-11}(L_{FIR}/L_\odot)$, which gives a SFR of 1 \msun/yr. \\
The H$\alpha$ line  is mainly emitted by HII regions where newborn massive stars are ionizing 
the gas. The total H$\alpha$ flux can be used as a tracer of the  star formation rate (Kennicutt 1998): \\
\begin{equation}
\mbox{L(H}\alpha) (erg/s) = 1.26\times10^{41} SFR (\msun/yr)
\end{equation}
The H$\alpha$ flux is 1.3 10$^{41}$ erg/s for the nuclear starburst for an aperture size of 3\arcsec 
(see  Table~\ref{tab_fluxes}), which gives a SFR of 1 \msun/yr, consistent with the SFR
derived from the FIR luminosity. These two derivations of the SFR take into
account only massive stars. Including all stellar masses, and
accounting for the effects of optical extinction, Eckart et al. (1991) derived 
a SFR of 4 \msun/yr for the nuclear starburst (NSB), and a supernova rate of $1.2\times 10^{-2}$/yr.

\subsubsection{(Non-)Thermal radiocontinuum}


\begin{table}
\caption{Outflow bubble and nuclear contributions to the H$\alpha$ and 20~cm 
radio continuum emission derived by fitting a 2D Gaussian to the central 
emission. The molecular gas mass for the NSB is the mass encompassed by one 
\twelvecoonezero beam, and has a projected size of 310$\times$250 pc.}
\label{tab_fluxes}
\begin{tabular}{lllll}
\hline
  & H$\alpha$ &  20 cm & 3.5 cm & M(H$_2$) \\
   & (10$^{39}$ erg.s$^{-1}$) &  (mJy) & (mJy) & (\msun)\\
\hline
\hline
  Nucleus & $\sim 130$  & $90$  & $28$ & $4.3\times10^8$ \\
  Outflow &   $\sim 45$   &  45 &  12  & $1.7\times10^7$\\
\hline
\end{tabular}
\end{table}

While the radio continuum emission at long wavelengths is mainly produced by the compact source (AGN) 
the contribution from the NSB is important at 3.5~cm  creating a flatter spectrum. 
By fitting the radio continuum between 6 cm and 80 cm 
where the non-thermal contribution is dominant, we found a spectral index of 
$\alpha=-0.46$. Extrapolating that fit down to 3.5~cm,
we obtained a non-thermal flux density of 6 mJy, which is much below the measured value 
of 31 mJy.

\begin{figure}
\resizebox{7.5cm}{5.5cm}{\includegraphics{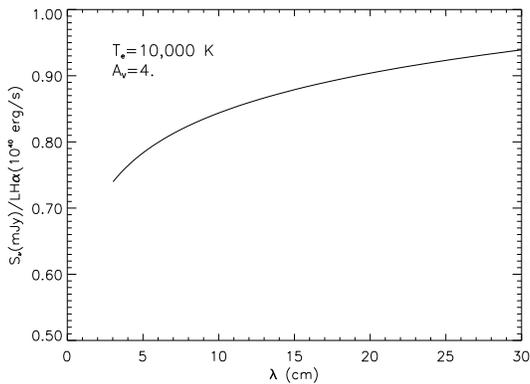}}
\caption{Ratio of the thermal radio continuum emission (mJy) and H$\alpha$ 
luminosity ($10^{40}$ erg/s) at the distance of NGC~6764, assuming an electron 
temperature of 10,000 K and an optical extinction of $A_V= 4$.
}
\label{fig_ratio_radio_halpha}
\end{figure}

The free-free contribution (``bremsstrahlung'') can be computed from the 
H$\alpha$ flux. The Balmer lines provide an estimate of the number of ionizing 
photons present in the source. If the optical extinction is known, the
H$\alpha$ flux can be used to compute the thermal radio continuum flux 
(e.g., Caplan \& Deharveng 1986).  Assuming an electron temperature of $10^4$ K 
and an optical extinction of $A_V= 4$
(Eckart et al. 1991), the thermal radio continuum flux density at 
3.5~cm is 14.8~mJy, obtained by integrating the H$\alpha$ luminosity in the 
outflow. The H$\alpha$ extinction is given by $A_{H\alpha} = 0.605*A_{V}$.  
Figure~\ref{fig_ratio_radio_halpha} shows the ratio between the radio flux 
density (mJy) and the H$\alpha$ luminosity (in units of $10^{40}$ erg/s) for
the thermal (free--free) contribution versus wavelength. This ratio is 
75\% at 3.5~cm.  The maximum thermal contribution to the flux density at 3.5~cm is 
thus about 23~mJy (Zurita et al. 2000) which is consistent with our previous 
estimation of 6~mJy for the non-thermal radiocontinuum.

Table~\ref{tab_radio_th-noth} gives an estimate of 
the thermal and non-thermal contributions to the observed radio continuum 
emission, using the H$\alpha$ luminosities to estimate the thermal component. At
20~cm the radio continuum emission is mainly non-thermal (87~\%), but at shorter 
wavelengths (6 and 3.5~cm) both contributions are comparable.

\begin{table}
\caption{Thermal (free-free) and non-thermal (synchrotron) contribution to the 
observed radio continuum emission at different wavelengths at the center of 
NGC~6764. The thermal contribution is estimated from the H$\alpha$ luminosity 
(see text).}
\begin{tabular}{llll}
\hline
Wavelength  & Thermal  & Non-thermal  & ratio $\frac{\mbox{Thermal}}{\mbox{Total}}$  \\
	   &  (mJy)	& (mJy)	    &  \\
\hline
\hline
20 cm	   &  17.6    & 115.4	     &  13\% \\
6 cm	   &  15.6    & 18.4	     &  46\% \\
3.5 cm	   &  14.8    & 16.3	     &  48\%  \\
\hline
\end{tabular}
\label{tab_radio_th-noth}
\end{table}

\section{Molecular gas}

\subsection{Distribution}

The CO emission detected with the PdBI is strongly concentrated toward the center of NGC~6764 (see 
Figure~\ref{fig_ico_pdbi}) and reaches a maximum at a projected distance of 
1\farcs 25 ($\sim$ 190 pc) from the radio continuum peak.
More than 98\% of the total molecular gas is concentrated
within the inner 800 pc, and as much as 
20\% of the total \twelvecoonezero emission is included within one beam.

\begin{figure*}
\resizebox{14cm}{16cm}{\includegraphics{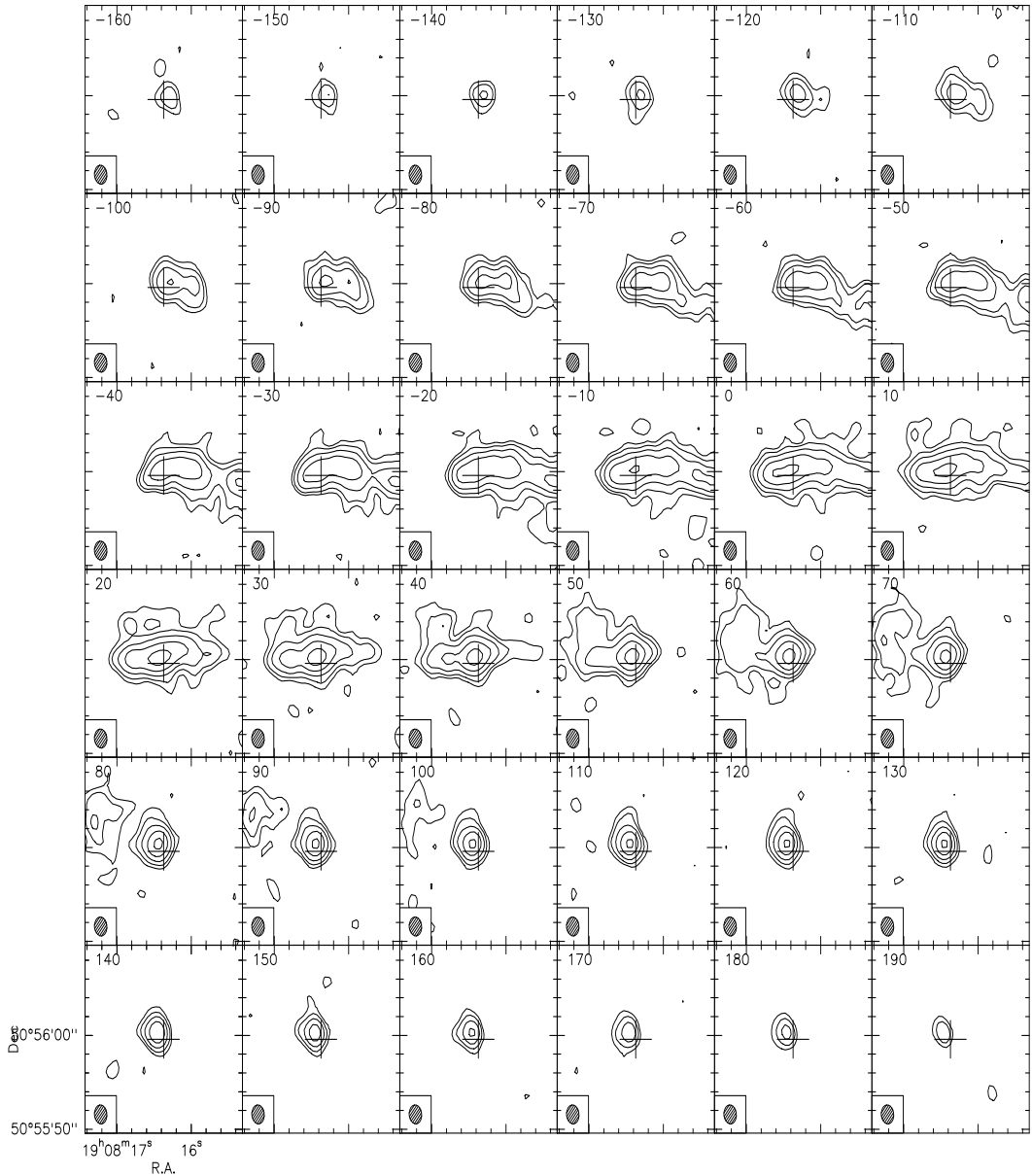}}
\caption{Channel maps for the \twelvecoonezero emission. The rms noise 
is 2 mJy~beam$^{-1}$. The contours are at (3, 6, 12, 24, 48, 92) $\sigma$. The synthesized 
beam is $2\arcsec\times 1\farcs 6$ in size. The rest velocity is 2400 \kms. The cross is located at 
the phase center, i.e., at $\alpha_{J2000}=$19:08:16.370 and $\delta_{J2000}$=50:55:59.58.}
\label{fig_co10_chanmap}
\end{figure*}

\begin{figure}
\resizebox{7.5cm}{7cm}{\includegraphics{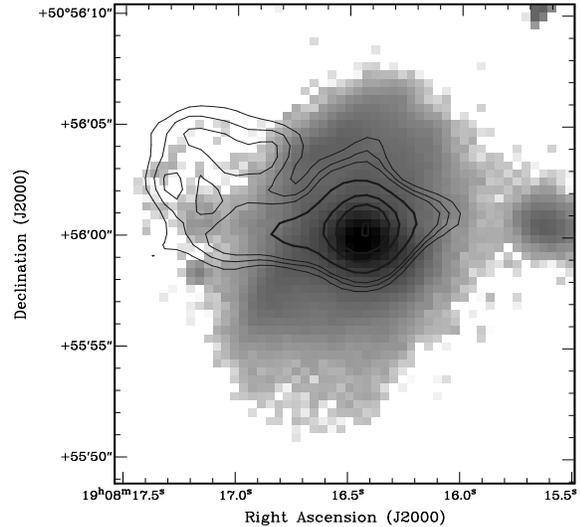}}
\caption{Integrated \twelvecoonezero (Jy.beam$^{-1}$\kms) emission intensity in the 
velocity range 20--100~\kms overlaid on the H$\alpha$ emission (greyscale). The contour levels are 
0.5, 0.75, 1, 2, 4, 6 and 10 Jy.beam$^{-1}$\kms (1$\sigma$ rms noise $\sim 0.02$ Jy.beam$^{-1}$\kms).}
\label{fig_Ico10_20_100}
\end{figure}


Apart from the central concentration, the CO 
emission is extended along the bar towards the western side (``W'') and 
has a smaller
extension to the eastern side. Assuming that the external spiral arms are 
trailing, this CO emission is on the leading side of the bar.

The western CO feature, which extends to 10\arcsec\ from the center, is aligned with
the primary bar. 
The CO emission is associated with two H$\alpha$-emitting star forming regions 
(with $5\times 10^{39}$ and 
$3\times 10^{39}$ erg.s$^{-1}$ H$\alpha$ luminosities not corrected for extinction) which are among 
the brightest in NGC~6764, excluding the strong H$\alpha$ emission at the tip 
of the bar (Rozas et al. 1996). Since NGC~6764 is highly inclined, 
the location of star formation and molecular gas in these regions is not 
completely clear. 

Figure~\ref{fig_co10_chanmap} shows the \twelvecoonezero channel maps at 
10 \kms\ velocity resolution relative to systemic. \twelvecoonezero emission is detected in the 
individual channels at 2.5$\sigma$ level from 2210 to 2600 \kms (LSR). Two 
\twelvecoonezero peaks are seen in the very center at negative and positive 
velocities. The first of these strong peaks is seen mainly at negative 
velocities ($-110$ to $-20$ \kms), and is separated from the primary CO peak by 
$\sim$2\arcsec. This secondary peak is not visible in the integrated 
\twelvecoonezero map. Extended weaker features are visible in the channel maps
as well. One feature is seen north of the \twelvecoonezero peak at velocities 
$-40$ to 130 \kms. Another more extended feature (see 
Figure~\ref{fig_Ico10_20_100}) is seen at velocities 20 to 100 \kms. The first 
feature is associated with the molecular outflow, visible in the 
\twelvecoonezero  integrated intensity image, and flaring up to a 8\arcsec 
($\sim$~1~kpc) height. Correcting for the disk inclination, the molecular gas reaches 
1.5 kpc above the plane of the galaxy.
The velocities are consistent with motion 
perpendicular to the galaxy plane, taking into account the inclination and 
orientation of NGC~6764.

The faint extended feature between 20 and 100 \kms\ covers an area of 
about 1 kpc$^2$ between 5\arcsec and 10\arcsec\ radii (800~--~1600 pc). At a 
velocity of 70 \kms, it reaches a total length of 12\arcsec\ ($\sim$1.9 kpc), 
and is oriented north to south. The velocity field shows few signs of 
outflow motion in this region, indicating that this feature is in the plane of
the galaxy. The western feature (W) has a large velocity extent from $-80$~\kms\ 
up to 10 \kms. A CO spur is visible towards the north, and its location is 
spatially correlated with the boundary of the radio continuum superbubble, as 
seen at 20 cm. 

\subsection{Kinematics}

\subsubsection{First and second velocity moments}
 
\begin{figure}
\resizebox{9cm}{8cm}{\includegraphics{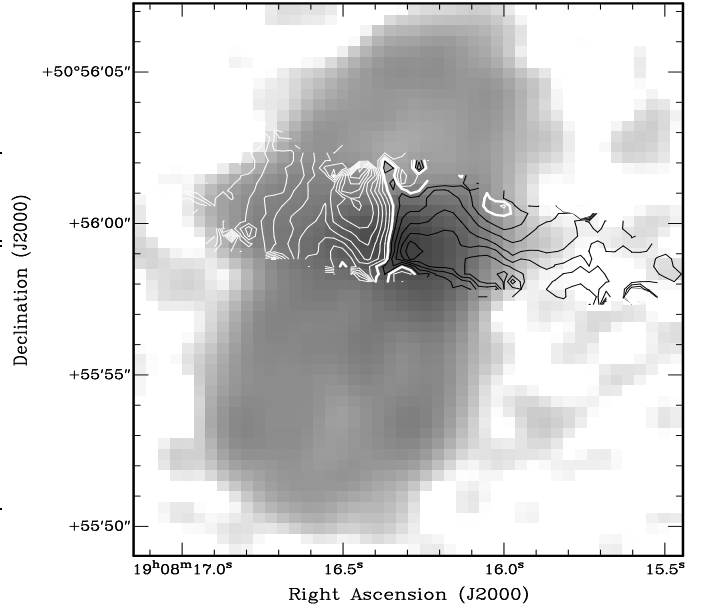}}
\caption{Isovelocity contours of the \twelvecotwoone emission overlaid on a
color image of the 20 cm radio continuum emission. Black contour lines indicate 
negative velocities, and are separated by steps of 10~\kms. White contour lines
imply positive velocities with respect to the reference velocity, and are also
separated by steps of 10~\kms. The reference velocity is 2400~\kms. 
}
\label{fig_co_mom1}
\end{figure}

Figure~\ref{fig_co_mom1} shows the beam-smeared map of the intensity-weighted 
\twelvecotwoone velocity field, derived from the first moment of the 
\twelvecotwoone data cube, overlaid on the 20 cm radio continuum emission. 
We determined the dynamical center of the gas motion from the \twelvecotwoone data,
and compared it to the positions of the various emission peaks (see 
Table~\ref{tab_astrom_peak}). \HS\ find an offset of 0.55\arcsec between the VLA
A-array peak at 3.5 cm and the optical nucleus. They explain that difference by a higher
extinction close to the nucleus, which is consistent with the molecular gas distribution.
The dynamical center of CO emission is spatially 
coincident with the nucleus as seen in the $B$- and $I$-bands, and located about
0\farcs 5 ($\sim80$ parsecs) from the radio continuum peak. Our observations 
are consistent with the global rotation of the galaxy disk, combined with 
streaming motions along the bar and/or a warping of the disk. The strongest 
deviation from the rotation pattern, the so-called ``spider diagram,'' is seen
on the eastern side along the molecular outflow, which originates from the 
\twelvecoonezero peak, at 1\arcsec--2\arcsec\ to the NE of the radio 
continuum peak. The noise in the velocity field there is significant since the CO 
emission is centrally concentrated and the emission intensity is much lower
beyond the peak. The position angle (P.A.) of the galaxy major axis, determined
from the velocity field, 70\degree$\pm 5$, is slightly larger than 
the P.A. from the orientation of the optical isophotes (61\fdg 5). 
The HI data of Wilcots et al. (2001) indicate a change in the
P.A. between the outer part  ($\sim 58\degree$) and the inner part of the disk 
($\sim 76 \degree$). 
The difference is most likely due to the streaming motion near the center caused by the bar,
as confirmed  by the disturbed CO velocity field. 

\begin{table}
\caption{Position (J2000.0) of the peak emission in radio continuum
(20 cm), CO, and H$\alpha$ (off-filter and the emission line). The
distance to the radio continuum peak is given in arcseconds.}
\label{tab_astrom_peak}
\begin{tabular}{lllll}
\hline
Tracer &  $\alpha_{2000}$ & $\delta_{2000}$  & D$_{20}$ & Astrom. \\
 &   &    &              & accuracy \\
&    &    &   (\arcsec)  & (\arcsec) \\
\hline
\hline
20cm    & 19:08:16.316 & 50:55:59.37 &  & 0.1 \\
\twelvecoonezero$_{peak}$ & 19:08:16.392 & 50:56:0.24 &  1.2  & 0.3 \\
\twelvecoonezero$_{vel. disp}$ & 19:08:16.392 & 50:55:59.88 &  0.9   & 0.3 \\
\twelvecotwoone$_{dyn}$ & 19:08:16.351 &  50:55:59.74  & 0.5 & 0.2  \\
\twelvecotwoone$_{peak}$ & 19:08:16.402 & 50:56:0.29  & 1.2  & 0.2 \\
H$\alpha$   & 19:08:16.440   &   50:55:59.87 &    1.3   & 0.2 \\
H$\alpha_{off}$ & 19:08:16.362 &  50:55:59.68  &  0.5   & 0.2  \\
Nucleus $B$-band &  19:08 :16.379 & 50:55:59.82 &	  0.6	 & 0.3  \\
Nucleus $I$-band &  19:08:16.374 &  50:55:59.69  &  0.6	 & 0.3 \\
\hline
\end{tabular}
\end{table}

\begin{figure}
\resizebox{8cm}{6cm}{\includegraphics{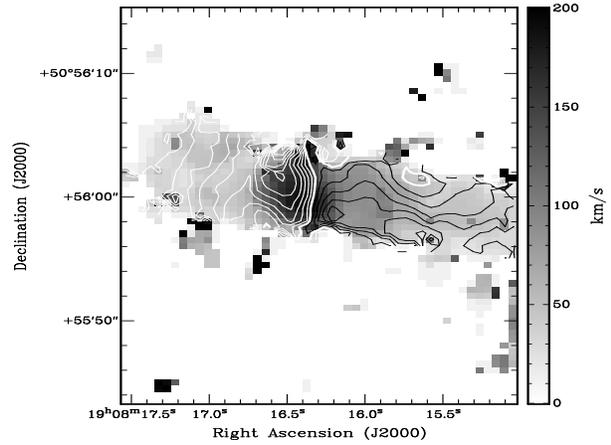}}
\caption{Velocity dispersion of the \twelvecoonezero emission in  greyscale, 
overlaid by the \twelvecoonezero velocity field. The white and black  lines are 
for positive and negative velocity relative to the reference velocity at 2400 \kms.}
\label{fig_co_mom2}
\end{figure}

Figure~\ref{fig_co_mom2} shows the beam-smeared map of the intensity-weighted 
\twelvecoonezero velocity dispersion in the center, derived from the second 
moment of the \twelvecoonezero data cube, overlaid with the 
\twelvecoonezero velocity field. The central velocity dispersion 
$\sigma_V$ = 200 \kms, is much higher than the internal velocity dispersion
of a molecular  cloud (typically a few  \kms; see e.g. Scoville et al. 1987). 
More likely, it results from the beam-smeared velocity 
gradient caused by a steeply rising rotation curve, together with a contribution
from streaming motions. 

Within the astrometric accuracy the peak of the velocity dispersion is 
located at the position of the dynamical center. 
At the position of the 
molecular outflow the velocity dispersion has values up to 150 \kms, whereas 
along the bar, at radii larger than 5\arcsec\ from the nucleus, the molecular 
gas velocity dispersion does not exceed 50 \kms. Curiously, the velocity
dispersion is asymmetric in the east--west direction, with larger values on the 
eastern side.

\subsubsection{Lindblad resonances}

\begin{figure}
\begin{center}
\resizebox{7.5cm}{5.5cm}{\includegraphics{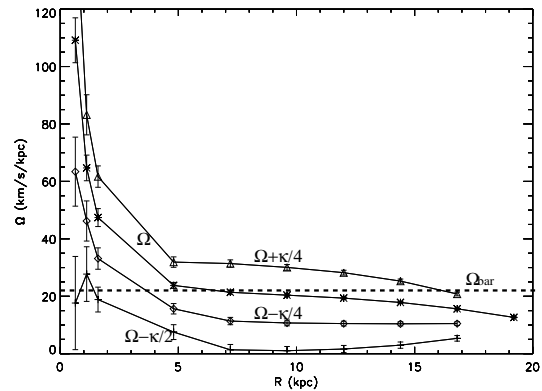}}
\caption{Azimuthal frequency $\Omega$ (stars) estimated from the rotation curve 
(HI+CO data). The first three data points are from our new CO (1--0) 
observations. The data points at larger radii are from HI observations 
(Wilcots et al. 2001). The UHR (diamond) and ILR (cross) locations have 
been estimated by setting the bar corotation radius to the semi-major axis radius of 
the bar ($\sim$40\arcsec). The bar pattern speed is estimated to be 22~\kmskpc.}
\label{fig_resonances}
\end{center}
\end{figure}

The rotation curve has been computed from our \twelvecoonezero data and the HI data 
from Wilcots et al. (2001), using the task INSPECTOR in {\tt Gipsy}. 
Although the HI spatial resolution is poorer than in CO, 
it provides data points at large radii. 
The rotation curve becomes flat at a value of 250 \kms\ at $\sim$ 12 kpc from the center. 
Under the assumption of the epicyclic approximation, the azimuthal frequency ($\Omega$) and the radial 
frequency ($\kappa$) have been computed from the rotation curve. The bar pattern speed is estimated 
from the rotation curve and the length of the bar semi-major axis
($\sim$6.5 kpc). Several studies (e.g., Athanassoula 1992, Combes \& Elmegreen 
1993, Sempere et al. 1995) have shown that the ratio of the bar corotation
radius to the bar semi-major axis length is in the range 1--1.5. The 
determination of the bar pattern speed is not strongly dependent on the exact 
value of this ratio since the frequency $\Omega$ is flat in that domain. The bar 
pattern speed $\Omega_{bar}$ is $\sim$22 \kmskpc\ with a corotation 
(CR) at a radius of 40\arcsec\ (6.5 kpc), using a value of 1 for the ratio of 
the radius of the CR to the length of the bar semi-major axis. The inner Lindblad
resonances (ILRs), if they exist, are the loci where the bar pattern speed 
equals $\Omega-\kappa /2$ (Figure~\ref{fig_resonances}).
Thus, the epicyclic approximation suggests that two ILRs are present near the 
center. The rough radii for the inner and outer ILRs (iILR and oILR) are
6\arcsec\ and 
10\arcsec\ (1 and 1.7 kpc). 4:1 resonances, or (inner/outer) ultraharmonic 
resonances (i/oUHR), with 
$\Omega_{bar} = \Omega\pm\kappa/4$, are found at radii of 24\arcsec\ 
(3.9 kpc; iUHR) and 100\arcsec\ (16.5 kpc; oUHR). At these resonances the star 
orbits are periodic and form rings (e.g, Byrd et al. 1998). Contrary to the UHR location, 
the locations of the ILRs strongly depend on the bar pattern speed. 
Table~\ref{tab_resonances} summarizes the locations of the
resonances found with the application of the epicyclic approximation, 
valid for asymptotically weak bars. In the presence
of an even moderately strong bar, an accurate determination of the resonance 
positions requires a full dynamical model, together with orbit analysis, 
which is beyond the scope of this paper.

\begin{table}
\caption{Dynamical resonances as estimated from a rough epicyclic model for
NGC~6764. $\Omega_{bar}$=22 \kmskpc. See the text for a detailed
description.}
\label{tab_resonances}
\begin{tabular}{lll}
Resonance & n   & Radius  \\ 
 & ($\Omega_{bar} = \Omega + n\times\kappa$) & \\
\hline
iILR  & -1/2 &  6$\pm 6$\arcsec \\
oILR  & -1/2 &  10$\pm 10$\arcsec \\
iUHR  & -1/4 &  24$\pm 5$\arcsec \\
CR    & 0  & 40$\pm 12$\arcsec \\
oUHR & +1/4 & 100$\pm 10$\arcsec \\
\end{tabular}
\end{table}

\subsubsection{Non-circular motions}

\begin{figure*}
\begin{center}
\resizebox{14cm}{18cm}{\includegraphics{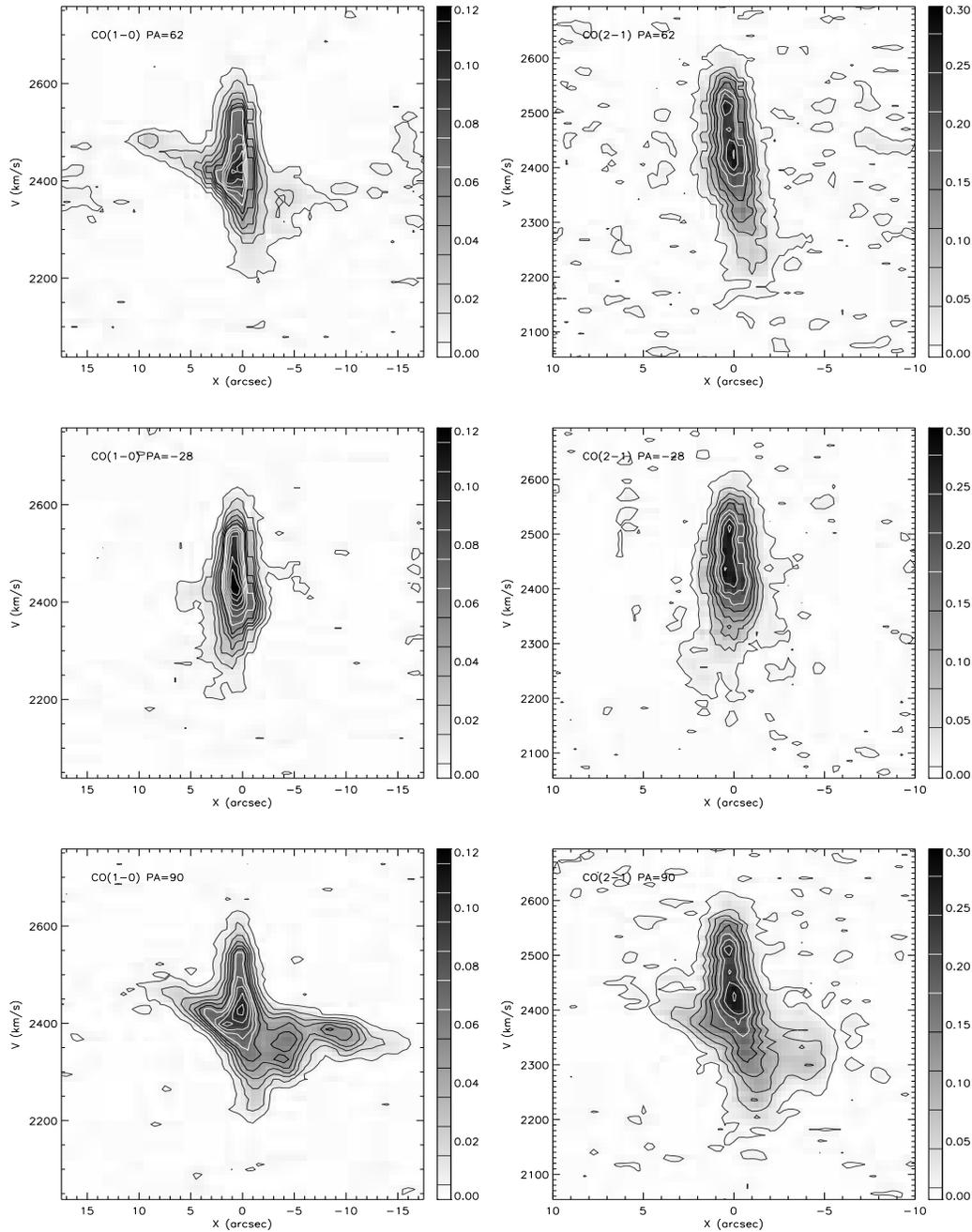}}
\caption{Position--velocity diagram of the \twelvecoonezero (left) and 
\twelvecotwoone (right) emission in the center of NGC~6764. The zero-offset is 
at the CO peak. The major axis P.A. is 62\degree, and the P.A. of 90\degree\ 
has been included to compare with Fig.~1 of Eckart et al. (1996). The units 
are in \jyperbeam.}
\label{fig_co_pvdiag}
\end{center}
\end{figure*}

Figure~\ref{fig_co_pvdiag} shows the position--velocity (PV) diagrams of the \twelvecoonezero and 
\twelvecotwoone emission along the major and minor axes (P.A.~=~62\degree\ and P.A.~=~152\degree)
of NGC~6764, along with a PV diagram at P.A.~=~90\degree\ for comparison with the
\twelvecoonezero observations made with the Nobeyama Millimeter Array (Eckart et al. 1996). 
As already shown by Eckart et al. (1996), the PV diagrams imply 
an asymmetric distribution of molecular gas near the nucleus.
The CO intensity is higher for gas at redshifted velocities, with respect
to the systemic velocity, i.e., on the eastern side. The PV diagram  at P.A. = 90\degree\ 
in the \twelvecoonezero emission shows strong streaming motion at 
x=$-4$\arcsec, which is located slightly inside the iILR ($r\sim6\arcsec$).
The \twelvecotwoone PV diagram along the minor axis reveals large 
streaming motions at x~=~1\arcsec--2\arcsec, with a velocity near $-200$ \kms. Similarly,
the \twelvecoonezero PV diagram along the minor axis shows an area of lower density gas on the 
eastern side at radii $<5\arcsec$, with streaming motions at negative 
velocities. The spur visible between x-positions 1\arcsec\ and 6\arcsec\ in the
velocity range [-30,80]\kms\ (relative to 2400~\kms) is located in the 
molecular outflow, and is consistent with the \twelvecoonezero spectrum integrated over the 
outflow area.
Assuming a trailing spiral 
structure, the northern part of NGC~6764 is closer to us, which implies
largely redshifted (positive) velocities for the radial outflow in the northern
side, as is seen in the PV diagram and in the integrated \twelvecotwoone 
spectrum.

\subsection{CO line ratio}

\begin{figure}
\resizebox{8cm}{7cm}{\includegraphics{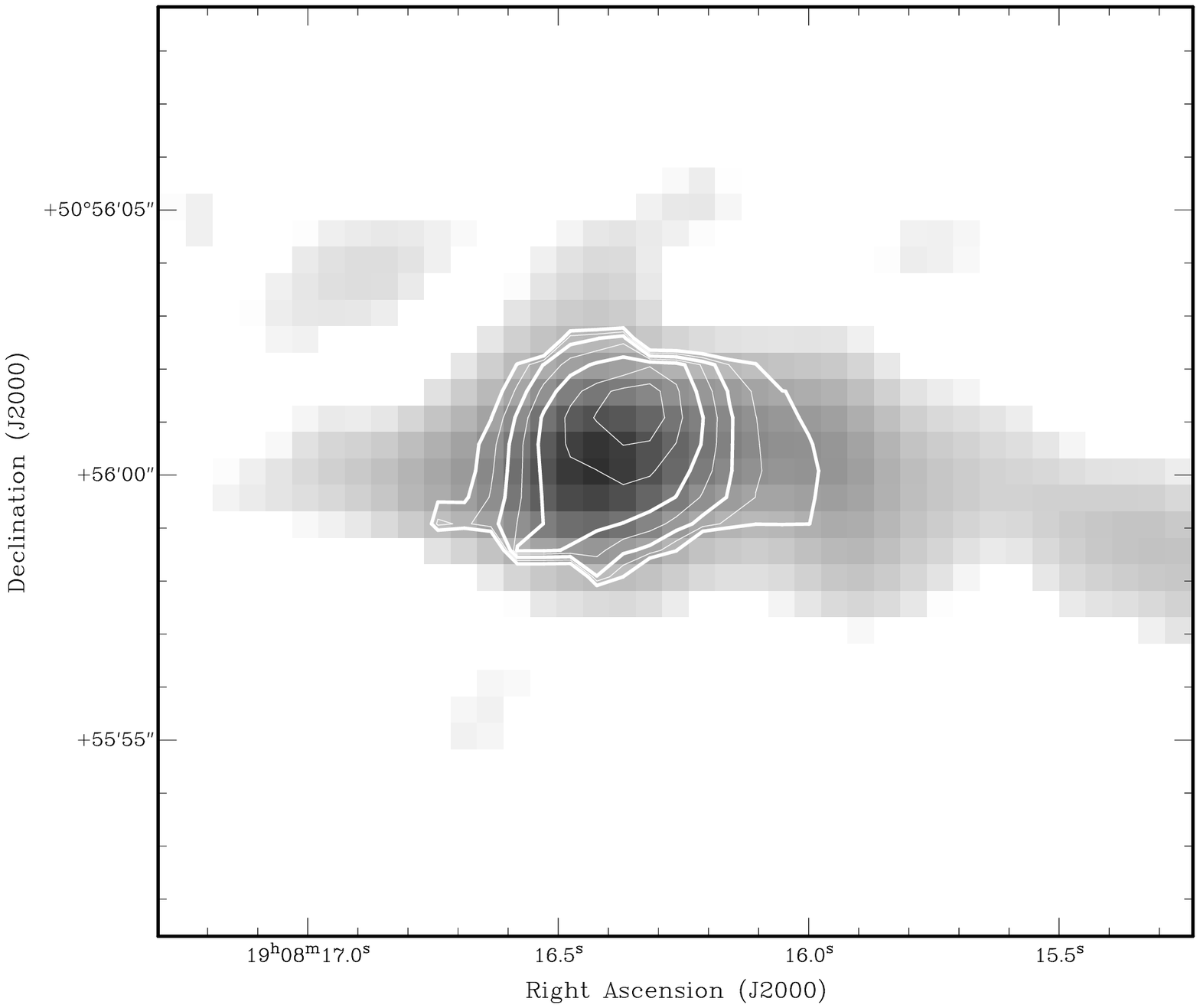}}
\caption{Integrated intensity \twelvecoonezero\ line (greyscale) overlaid by contours
showing the \twelvecotwoone\ to \twelvecoonezero\ line ratio, integrated over all 
channels. The contours are 0.4, 0.5, 0.6, 0.7, 0.8, 0.9 and 0.95.}
\label{fig_co_line_ratio}
\end{figure}

The line ratio for the integrated CO intensity between the \twelvecotwoone\ and 
\twelvecoonezero\ transitions,  
$\Re = I(\twelvecotwoone)/I(\twelvecoonezero)$, was computed matching the u-v coverage of both transitions 
and using the same spatial resolution and taking only  the \twelvecoonezero and \twelvecotwoone data above a threshold of $3\sigma$.
The result is shown in Figure~\ref{fig_co_line_ratio}. 
The maximum $\Re$ is 0.98, in agreement with that estimated
by Eckart et al. (1991) from single dish observations. Eckart et al. (1991) interpreted 
the global ratio to arise from moderately warm, dense, and optically thick gas.

\begin{figure}
\resizebox{7.5cm}{5.5cm}{\includegraphics{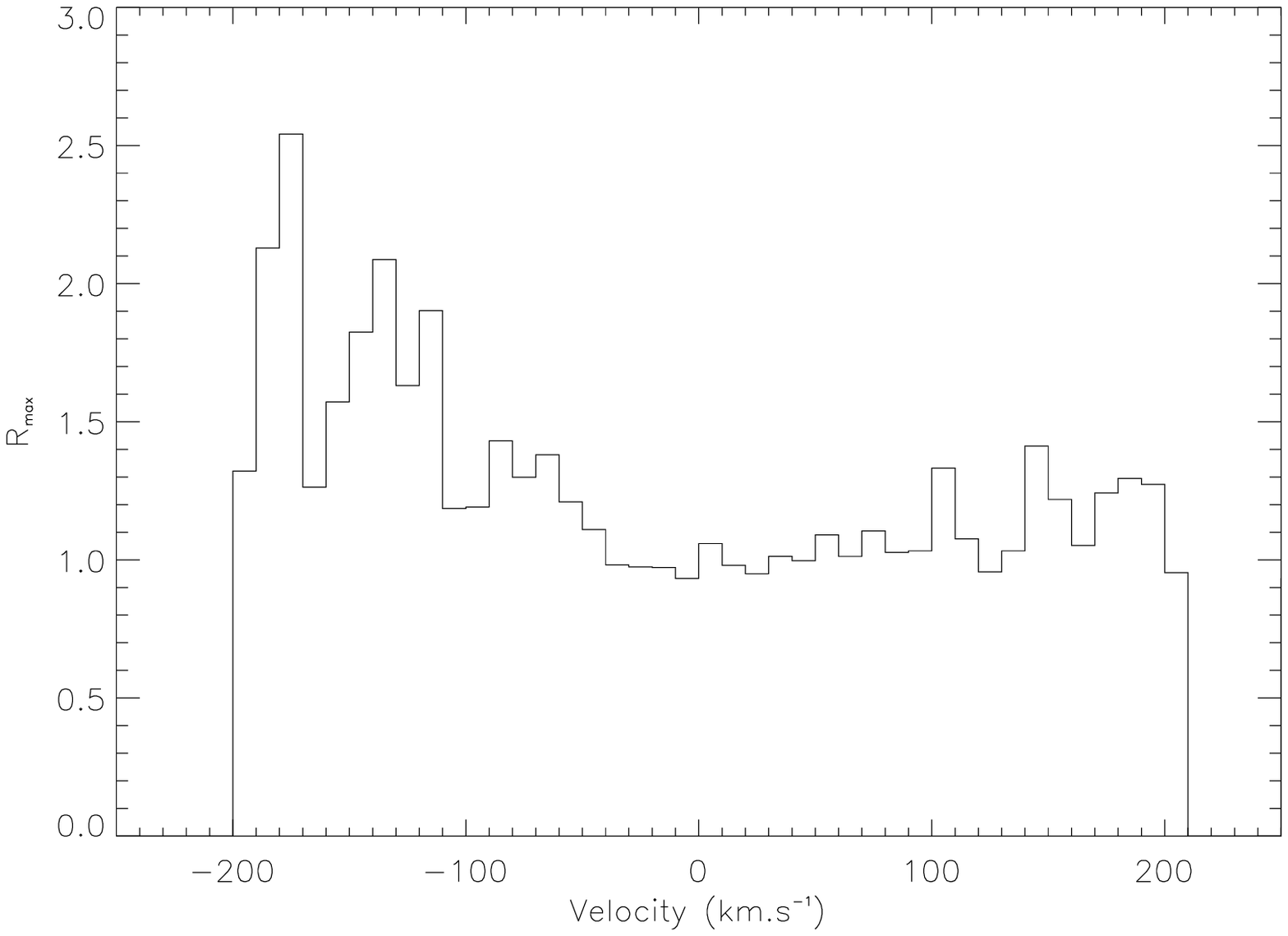}}
\caption{Maximum \twelvecotwoone\ to \twelvecoonezero\ line ratio $\Re_{max}$ 
computed in the center of NGC~6764 for each \twelvecoonezero\ channel 
(10 \kms width).}
\label{fig_coratiomax_vel}
\end{figure}

\begin{figure*}
\begin{center}
\resizebox{15cm}{20cm}{\includegraphics{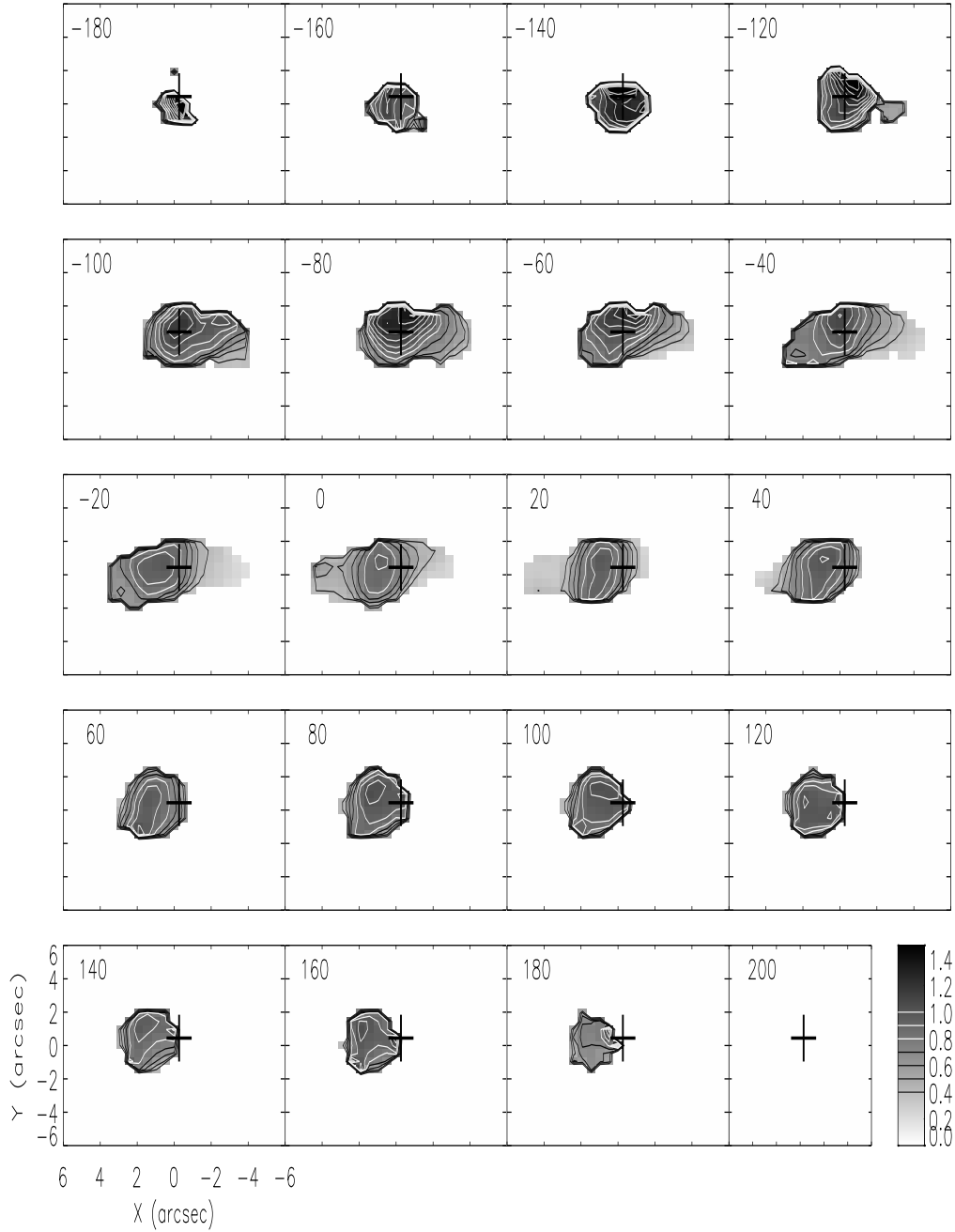}}
\end{center}
\caption{\twelvecotwoone to \twelvecoonezero line ratio for different 
velocities, as indicated in each panel, with a channel width of 10 \kms. The 
first contour is at 0.4 and the contours increase in steps of 0.1. The radio 
continuum peak is shown by a cross.}
\label{fig_co_line_ratio_chan}
\end{figure*}

To estimate the uncertainty in the line 
ratio introduced by the missing flux, we compared the CO line ratio map 
derived from the single-dish data with the ratio derived from 
our new PdBI observations.
We derive larger values of the ratio at large radii, owing to the lower 
spatial resolution ($\sim 20 \arcsec$) of Eckart et al. (1991). Their maximum
value at the center is lower than ours by $\sim 10$\%.  
The maximum line ratio $\Re_{max}$ in each channel (10 \kms~width)
is shown in Figure~\ref{fig_coratiomax_vel}. A bimodal feature is clearly 
seen, with peaks towards the blue and red sides. The blue side with a peak at 
$-140$ \kms\ is associated with molecular gas closer to the radio continuum 
peak, as shown in the CO line ratio channel maps in 
Figure~\ref{fig_co_line_ratio_chan}. At the peak the ratio reaches values of
2--2.5.
The red side of the peak at 140 \kms\ has a maximum value of about 1.3. This high
CO line ratio in fact applies to the bulk of the molecular gas, as is shown 
by the CO line ratio channel maps below.

The \twelvecotwoone/\twelvecoonezero line ratio has been computed for each 
channel (10~\kms width). Figure~\ref{fig_co_line_ratio_chan} shows the CO line 
ratio from a channel at $-180$ \kms\ up to a channel at 200 \kms\ (the systemic
velocity of NGC~6764 is 2400 \kms, corresponding to a zero velocity in the 
channel maps). From $-180$ 
to $-40$~\kms ~the peak of the CO line ratio is spatially coincident with the 
nucleus (or the radio continuum peak, as indicated by a cross in the figure). 
The peak at $-140$ \kms\ seen in Figure~\ref{fig_coratiomax_vel} is likely 
to be an edge effect. The ``true'' maximum is reached at $-130$ \kms, where the
ratio reaches a value of 1.4 at the exact position of the radio continuum peak 
(to within a pixel width). The western CO emission feature has a mean CO 
line ratio of 0.3, which is likely to be an underestimate because of the 
filtering out of the extended emission.

The high CO line ratios for the blueshifted gas in the nucleus may be 
associated with outflowing molecular gas powered by the AGN or the starburst. \HS\
reported a possible blueshifted HI absorption at $\sim 120 \kms$ against the compact core which may
be related to the molecular gas with such high CO line ratios.
We note that the radio continuum map (Fig.~\ref{fig_vla20_35}) has hints of a nuclear radio jet 
towards the south, likely associated with blue-shifted motions there as expected
if the orientation of NGC~6764 with respect to the line of sight is taken into
account. A redshifted counterpart with such a high line ratio is not detected.

\subsection{\htwo\ mass}

Adopting a standard conversion factor of 
X$_{\mbox{CO}} = 2.3\times10^{20}~$cm$^{-2}$ (K~km~s$^{-1}$)$^{-1}$ (e.g., 
Strong et al. 1988), the total mass of molecular hydrogen (\htwo) is:

\begin{equation}
M_{H_2} = 9~\times~10^3~S_{CO}~D^2
\end{equation}

\noindent where $S_{CO}$ is the integrated CO flux density in Jy~\kms\ and $D$ 
is the distance in Mpc. 
The integrated \twelvecoonezero\ flux density from our new map gives a total 
molecular hydrogen mass of $1.0\times10^9$~\msun. It does not take into account
the 50 \% of flux filtered out by the interferometer observations.

In the Milky Way's disk the X$_{\mbox{CO}}$ conversion factor derived from
observations
has been found to match the theoretical conversion factor quite well (e.g., 
Radford et al. 1991). However, there is growing evidence that the
conversion factor is overestimated by a factor of 3--10 in the central part of 
galaxies, especially in those with nuclear starbursts (Downes \& Solomon 1998). The physical reasons for 
this discrepancy have been discussed in several papers, and include 
the metallicity gradient (e.g., Dahmen et al. 1998), broadening of the CO line by 
the bulge potential (e.g., Stacy et al. 1989, Mauersberger et al. 1996), 
and changes in the physical conditions of the circumnuclear gas (e.g., its optical 
depth, temperature, and density). In NGC 253, a nearby starburst galaxy, 
Mauersberger et al. (1996) found a conversion factor $\sim 7$ times lower than 
$X^{std}_{CO}$ in the central part of the galaxy. They argued that the 
central potential well is broadening the CO lines, and its emission could be 
partially optically thin. The conversion factor has been found to be 3 to 10 
times lower in the galactic bulges compared to their disks (Sodroski et al. 
1995). Recently, Wei\ss\ et al. (2001) analyzed the variations of X$_{\mbox{CO}}$
in the starburst of M82 in detail. They found that `` X$_{\mbox{CO}}$ is a 
function of the intrinsic gas properties, which strongly depend on 
environmental effects'' (massive star formation). This leads to a variation of 
X by a factor of 5--10, compared to the Galactic value of 
X$_{\mbox{CO}}= 1.6\times10^{20}$~cm$^{-2}$ K~km~s$^{-1}$. 

 The dust mass 
computed from the IRAS fluxes at 60 and 100 \micron\ gives a total mass of 
M$_d$=$1.4\times10^6\msun$ with a mean temperature of 36.5~K. It is known from 
FIR observations with ISO that the IRAS data are missing a cold dust component 
that emits beyond 150 \micron\ and has a temperature of T$_d < 20$ K, 
(Alton et al. 1998, Trewhella et al. 2000). This component can be 
2--10 times more massive than the warm dust component. 

In NGC~6764 the gas-to-dust mass ratio, using the standard CO-to-H$_2$ conversion 
factor and correcting for the missing flux in the interferometer map, 
leads to $M_{\htwo}/M_{dust} \sim 1500\pm 225$. This factor is much larger than 
the typical value found in normal spiral galaxies (an average ratio of 
$\sim 700$; Leon et al. 1998) using IRAS-derived dust masses. This indicates that
the molecular gas mass obtained from the \twelvecoonezero luminosity and the 
standard conversion factor has most likely been overestimated.

The high gas-to-dust mass ratio  means that the 
\htwo\ mass has probably been overestimated by a factor of 2--3 in the center of 
NGC~6764, when using the standard conversion factor X$\sim X^{std}_{CO}$.
This overestimate may be even more severe in the nucleus because of the intense 
starburst. 
Thus we estimate the total \htwo\ mass to be 7$\times 10^8\msun$,
taking into account the filtered out flux in the \twelvecoonezero interferometric map.

\section{Outflow}

The basic morphological parameters for the outflow
from the 20 cm radio continuum observations are: a projected base of radius 3.5\arcsec (540 pc) and a projected height 
of 7\arcsec (1080 pc) above the galaxy plane. The mass-load in the outflow is 
estimated from the spatial distribution of the \twelvecoonezero emission. The \twelvecoonezero emission was integrated in the area
shown in  Fig. \ref{fig_ico_pdbi} only for the northern part of the outflow. The area was chosen in order 
to avoid the contamination from the molecular gas emission in the disk, given the inclination of the galaxy and the spatial
resolution of the observations. The  \twelvecoonezero integrated intensity in the out-of-plane outflow is  1.4 Jy~\kms.

The observed energy of the different gas phases in the outflow is an indicator 
of the physical processes responsible for the outflow. 
Assuming that the bulk of the molecular gas is
perpendicular to the disk plane, the expansion velocity V$_{exp}$ is estimated from the spur in \twelvecoonezero seen 
between -30 and 80 \kms  on the PV diagram (see Fig. ). We take a projected expansion  velocity of 25 \kms, which gives 
 an expansion velocity of 53 \kms. This value is quite uncertain since we know already that some  molecular gas is reaching 
an expansion velocity of 170 \kms.
Applying a correction of 1/3 to the CO-to-H$_2$ conversion factor (see \S 4.4) 
the total molecular gas mass in the outflow is $4.3\times 10^6$ \msun. Thus the 
molecular gas kinetic energy E$_{kin}$ in the outflow is 
$E_{kin}=\frac{1}{2} M_{gas} * V_{exp}^2=2.4\times 10^{53}$ ergs. 


For a starburst with a SFR=1~\msun/yr, different 
star formation models have been computed (instantaneous or continuous). The 
total energy released by SNs and stellar wind at the age of the recent NGC~6764 
starburst ($\sim$ 5 Myr) is in the range of $10^{54}-10^{55}$ ergs. 
Thus there is no need to invoke AGN or other mechanisms, such as hypernovae or tidal 
interaction, to drive the outflow, even if the AGN seems to have an effect on the molecular
gas physical conditions (see Sect. 4.3) and .the data cannot exclude  the AGN  from being 
the major energy source either.

To further study the energy budget, we computed the total thermal energy 
$E_x$ in the hot X-ray gas, which was determined by the HRI {\em ROSAT} 
instrument. The X-ray emission is composed by two components: one varying on the timescale 
of several days and a more diffuse one which  extends up to 15\arcsec\ from the
nucleus, roughly along the minor axis (Schinnerer et al. 2000). We make
use of the X-ray parameters derived by Eckart et al. (1996) from the {\em ROSAT} 
observations in the 0.1--2.4 keV band. 
They estimated an electron density 
$n_e \sim 0.2f_s^{-1/2} \mbox{cm}^{-3}$, where $f_s$ is the volume filling 
factor, the volume of the extended X-ray gas as 
$V\sim 7\times10^{63} \mbox{cm}^{3}$, and the temperature of the X-ray gas as
$T\sim 4\times10^6$ K. The total thermal energy is then given by 
$E_x \sim 3n_ekTVf_s$, i.e. $E_x \sim 2.3\times10^{54} f_s^{1/2}$ ergs. 
We use $f_s = 0.01$, as was found for the starburst galaxy NGC~3256 
(Heckman et al. 1990; Moran et al. 1999). 
Thus $E_x$ is estimated to  be $\sim 2.3\times10^{53}$ ergs. The current 
starburst in NGC~6764 appears to be capable of releasing sufficient energy to 
power the hot X-ray gas bubble.

To estimate the distribution of the radio continuum emission at 20 cm in the
outflow bubble, the nuclear component has been subtracted by fitting a Gaussian 
to the nuclear flux above a threshold of 4 \mjyperbeam. 
The total flux remaining in the center-subtracted map is 
63 mJy, or about 60\% of the total seen in the VLA observation.
There is an offset of the centroid of the
outflow emission relative to the position of the nuclear radio continuum peak, 
and relative to the disk plane. This offset can be due to various reasons: a) the starburst is located 
in the western part of the nucleus, and  slightly below the plane, as suggested by the lower gas column density south of
the nucleus, b) there is a possible contribution from a radio jet 
in the southern part in addition to the starburst-induced outflow.

\section{Discussion}

We compare the CO and  H$\alpha$ observations to the different 
hydrodynamical models of Tomisaka \& Ikeuchi (1988) performed for a smooth gas component with various
scenarios of starburst.
The case D of Tomisaka \& Ikeuchi (1988) has a starburst with a supernova
rate comparable to the one found in NGC~6764 ($10^{-2}$yr$^{-1}$: the bubble reaches z$\sim$1.2 kpc after 9 Myr,
which is difficult to conciliate with the W-R star age and population. In the case C, with an ambient 
density of $n\approx 100$ $\mbox{cm}^{-3}$, the height of the bubble reaches 1 kpc after 7 Myr. Moreover
they performed a simulation with a short ``explosion''  which drives the bubble height to z $\approx$ 1.3
kpc as fast as in 1 Myr. From all these different models, we can conclude that the recent starburst 
has been intense and short 3-7 Myr ago, in complete agreement with the stellar population analysis of
Schinnerer et al. (2000). From the comparison with these simulations we can conclude that the hot gas in NGC~6764 is still 
in a confined bubble, and has not yet escaped to the halo.

The NSB in NGC~6764 is very similar to the NSB in NGC~2782 (Jogee et al. 1998, 
1999). However, NGC~2782 has no nuclear activity, apart from the starburst. The 
CO spurs in NGC~2782 have a total molecular gas mass of $2\times10^7$ \msun, 
a total molecular gas kinetic energy of $\sim 8 \times 10^{53}$ ergs, and
a thermal energy of $\sim 2\times10^{56}f^{0.5}$. These numbers are larger by  
factors of 5--10 compared to the energy released in NGC~6764. Nevertheless the 
size of the outflows is comparable. This is probably due to the lower ISM 
density in NGC~6764 than in NGC~2782.


Observations (Seaquist \& Clark 2001, Walter et al. 2002) of  \twelvecothreetwo and \twelvecoonezero emission  in the prototypical starburst 
galaxy M82 indicate a large amount of molecular gas ($> 3 \times 10^8 \msun$ in the outflow with a maximum outflow velocity of
230 \kms and an estimated kinetic energy of $\sim 10^{55} ergs$. Similarly to our estimate, Wei\ss\ et al. (2001) 
found a CO-to-\htwo\ conversion 3 times lower than the Galactic conversion factor. The height of the molecular gas 
outflow is about 1 kpc above the plane with an estimated starburst age of 10 Myr. The comparison with NGC~6764 indicates
that in this galaxy the build-up of the molecular outflow was faster by a factor 2-3. The main difference  is the
molecular gas mass involved in the outflow, which is an order of magnitude larger  in M82 than in NGC~6764.

\HS\ present a compilation of 10 galaxies with non-thermal bubbles that shows that all are associated with an AGN. They link  the origin 
of these bubbles to the presence of an AGN. We note that 6  galaxies of their sample (60 \%) have a FIR luminosity larger than the one of NGC~6764, if we adopt their luminosities. The presence of an AGN may not be  sufficient to create these bubbles.
It may indicate that the association of a  starburst, present or recent, with an AGN could be the key to generate non-thermal radio bubbles. 
The creation of such feature would depend on a short time scale combination of the two processes: the starburst and the AGN. Nevertheless we note
that the presence of radio AGN is very  low in spiral galaxies (J. Sabater, private communication) which makes the correlation between radio-AGN and
non-thermal bubble very significant. Thus a statistical study with a large sample of galaxy with non-thermal bubbles would help to disentangle their origins.

NGC~6764, like other composite galaxies with a nuclear starburst and an AGN, shows that the molecular gas evolution in the very center
is mainly driven by the starburst, especially in the molecular outflow. Even if the AGN has an active  role in the central evolution, as shown 
by the gas chemistry in the
Seyfert 2 galaxy NGC~1068 (Usero et al. 2004), the only possible influences of the AGN on the molecular gas, at the resolution of the observations of the CO emission in NGC~6764, appears to be the high CO line ratio in the very center, and likely  the high HCN/CO ratio
already found by Kohno et al. (2001). No clear dynamic signature
is observed. The AGN are able to inject vast amounts of energy into their host galaxies and should have
a disruptive influence on the molecular gas in the central regions. 

\section{Summary}

We have found the following results from our high-resolution mapping of the 
\twelvecoonezero\ and \twelvecotwoone\ millimetre lines and 3.5 and 20 cm radio 
continuum emission in the circumnuclear area of the NSB galaxy NGC~6764.

\begin{enumerate}

\item The bulk of the $^{12}$CO emission is concentrated in the very 
center of NGC~6764 and is offset from the 20 cm radio continuum peak 
by $\sim 1\farcs 2$ (with respect to the \twelvecoonezero emission peak). 
A more extended component is detected along the bar on the western side of
the nucleus. 

\item Non-thermal radio continuum emission  at 3.5 and 20 cm is observed above 
and below the galactic plane in a nuclear  bubble-like outflow, as already shown
by \HS\ and Laine et al. (2006). This 
outflow is detected out to z=1.3 kpc above the galactic plane (deprojected). 
\twelvecoonezero emission is detected at the NE boundary of the outflow. About 
$4.3\times 10^6~\msun$ of molecular gas is driven by the outflow out of the 
plane of the galaxy.

\item The total molecular gas mass is $6.7\times10^8 \msun$ using a CO-to-H$_2$
conversion factor equal to 1/3 of the Galactic value. The use of such a value 
is motivated by the large gas-to-dust mass ratio and previous studies of 
molecular gas in starburst galaxies.

\item The \twelvecotwoone / \twelvecoonezero line ratio was computed in each 
channel (10 \kms velocity width). It reaches a maximum of 1--2 at the nucleus,
and is likely associated with the nuclear AGN (it coincides with the 20 cm 
radio continuum peak). In the bulk of the molecular gas the maximum ratio is 
about 1--1.3, consistent with warm (T$_{kin} \ge 20~$K) and dense gas 
($n_{\htwo} > 2\times 10^4 \mbox{cm}^{-3}$), as found by Eckart et al. (1991). 


\item An important fraction of the radio continuum at 3.5 cm is thermal. The 
outflow morphology and the ISM properties indicate an age of $\sim$ 3--7 Myr 
for the recent starburst, by comparing the observations to hydrodynamical 
simulations. It appears that the outflow is still confined and not freely expanding.
A ``hot-spot'' of thermal gas with a flat spectral index ($\sim -0.2$) is 
located near the northern boundary of the bubble/outflow.

\item The  kinetic and thermal energies  of the molecular and 
hot gas components in the bubble can be fully accounted for by the energy 
released from the NSB, even if the data cannot exclude the AGN from being the 
major energy source.


\end{enumerate}


\begin{acknowledgements}
We thank an anonymous referee for his  careful reading and very  detailed report which helped 
 to improve this paper significantly.
This research has made use of the NASA/IPAC Extragalactic Database (NED) which 
is operated by the Jet Propulsion Laboratory, California Institute of 
Technology, under contract with the National Aeronautics and Space 
Administration. During  part of this research S. Leon was supported by a Marie Curie 
Individual Fellowship  contract HPMF-CT-20002-01734 from the European Union. We would like to thank 
David Frayer and Eric Wilcots  for providing their H$\alpha$  and HI data, respectively.
\end{acknowledgements}


\begin{thebibliography}{}

\bibitem{} Alton, P. B., Bianchi, S., Rand, R. J., Xilouris, E. M., Davies, J. I., Trewhella, M. 
1998, ApJ, 507, 125
\bibitem{} Athanassoula, E. 1992, MNRAS, 259, 345
\bibitem{} Baum, S. A., O'Dea, C. P., Dallacassa, D., de Bruyn, A. G., Pedlar, A. 1993 ApJ, 419, 553
\bibitem{} Byrd, G.G., Ousley, D., dalla Piazza, C. 1998, MNRAS, 298, 78
\bibitem{} Caplan, H., Deharveng, L. 1986, A\&A. 155, 297
\bibitem{} Colina, L., Arribas, S., Borne, K.D. 1999, ApJ, 527, L13
\bibitem{} Combes, F. 2001, 'The Central Kiloparsec of Starbursts and AGN: The La Palma Connection', 
ASP Vol. 249. Eds.  Knapen, J.H., Beckman, J.E., Shlosman,I, Mahoney, T.J, p. 475
\bibitem{} Combes, F., Gerin, M.  1985, A\&A, 150, 327 
\bibitem{} Combes, F., Elmegreen, B.G. 1993, A\&A, 271, 391 
\bibitem{} Condon, J. J., Yin, Q. F. 1990, ApJ, 357, 97 
\bibitem{} Dahmen, G., Huttemeister, S., Wilson, T. L., Mauersberger, R. 1998, A\&A, 31, 959
\bibitem{} Dopita, M.A. 1985, ApJ, 295, 5
\bibitem{} Downes, D., Solomon, P. M. 1998, ApJ, 507, 615
\bibitem{} Eckart, A.,  Cameron, M., Boller, Th., Krabbe, A., Blietz, M., Nakai, N., 
Wagner, S. J., Sternberg, A. 1996, ApJ, 472, 588
\bibitem{} Eckart, A., Cameron, M., Jackson, J. M., Genzel, R., Harris, A. I., Wild, W.,
 Zinnecker, H. 1991, ApJ, 372, 67
\bibitem{} Franceschini, A., Silva, L., Fasano, G., Granato, L., Bressan, A. Arnouts, S., Danese, L. 1998,
ApJ, 506, 600
\bibitem{} Hota, A, Saikia, D.J. 2006, MNRAS, 371, 945 (\HS)
\bibitem{} Heckman, T. M., Armus, L., Miley, G. K. 1990, ApJS, 74, 833
\bibitem{} Hummel, E., van Gorkom, J. H., Kotanyi, C. G. 1983, ApJ, 267, 5
\bibitem{} Hummel, E., Beck, R, Dettmar, R.-J. 1991, A\&AS, 87, 309
\bibitem{} Hill, T. L., Heisler, C. A., Norris, R. P., Reynolds, J. E.; Hunstead, R. W. 2001,
 AJ, 121, 128
\bibitem{} Irwin, J. A., Sofue, Y. 1996, ApJ, 464, 738
\bibitem{} Jogee, S. Kenney, J. D. P., Smith, B. J. 1998, ApJ, 494, 185
\bibitem{} Jogee, S., Kenney, J. D. P., Smith, B. J. 1999, ApJ, 526, 665
\bibitem{} Kenney, J. D. P., Carlstrom, J. E., Young, J. S. 1993, ApJ, 418, 687
\bibitem{} Kennicutt, R. C., Jr. 1998, ApJ, 498, 541
\bibitem{} Kohno, K. et al. 2001, ``The Central Kiloparsec of Starbursts and AGN: The La Palma Connection'', 
ASP Conference Proceedings Vol. 249. Eds. Knapen, J. H., J. E. Beckman, J.E.,  Shlosman, I. and Mahoney, T.J. 
San Francisco: Astronomical Society of the Pacific, 2001, p. 672.
\bibitem{} Kunth, D. Legrand, F. Tenorio-Tagle, G. Silich, S. Mas-Hesse, J. M. Cervi\~no, M. 2002, Ap\&SS, 
281, 261
\bibitem{} Krolik, J.H., Begelman, M.C. 1986, 308, L55
\bibitem{} Laine, S., Kotilainen, J. K., Reunanen, J., Ryder, S. D., Beck, R.
 2006, AJ, 131, 701
\bibitem{} Laine, S., Shlosman, I., Knapen, J. H., Peletier, R. F. 2002, ApJ, 567, 97 
\bibitem{} Launhardt, R., Zylka, R., Mezger, P. G. 2002, A\&A, 384, 112
\bibitem{} Lee, S.-W., Irwin, J. A., Dettmar, R.-J., Cunningham, C. T., Golla, G., Wang, Q. D. 2001, A\&A, 377, 759
\bibitem{} Leon, S., Combes, F., Menon, T. K. 1998, A\&A, 330, 37
\bibitem{} Leon, S., Meylan, G., Combes, F. 2000, A\&A, 359, 907
\bibitem{} Mauersberger, R., Henkel, C., Wielebinski, R., Wiklind, T., Reuter, H.-P. 1996, A\&A, 
305,.421
\bibitem{} Mihos, J. C., Hernquist, L. 1994, ApJ, 425, 13
\bibitem{} Moran, E. C., Lehnert, M. D., Helfand, D. J. 1999, ApJ, 526, 649
\bibitem{} Norman, C. A., Ikeuchi, S. 1989, ApJ, 345, 372
\bibitem{} Osterbrock, D. E., Cohen, R. E. 1982, ApJ, 261, 640
\bibitem{} Radford, S. J. E., Downes, D., Solomon, P. M. 1991, ApJ, 368L, 15
\bibitem{} Rozas, M.; Beckman, J. E.; Knapen, J. H. 1996, A\&A, 307, 735
\bibitem{} Sakamoto, K. Ho, P. T. P., Iono, D., Keto, E. R.; Mao, R.-Q., Matsushita, S., Peck, A. B., Wiedner, M. C.; Wilner, D. J., Zhao, J.-H. 2006, ApJ, 636, 685
\bibitem{} Sanders, D. B., Mirabel, I. F. 1985, ApJ, 298, 31
\bibitem{} Scalo, J., Chappell, D. 1999, MNRAS, 310, 1
\bibitem{} Scoville, N. Z., Yun, M. S., Sanders, D. B., Clemens, D. P., Waller, W. H. 1987, ApJS, 63, 821
\bibitem{} Schinnerer, E.,  Eckart, A., Boller, Th.  2000, ApJ, 545, 205
\bibitem{} Seaquist, E. R., Clark, J. 2001, ApJ, 552, 133
\bibitem{} Sempere, M.J., Garcia-Burillo, S., Combes, F., Knapen, J. H. 1995, A\&A, 296, 45
\bibitem{} Sodroski T.J. et al. 1995, ApJ, 452, 262
\bibitem{} Stacy, J. G., Bitran, M. E., Dame, T. M., Thaddeus, P. 1989 The Center of the Galaxy:
 Proceedings of the 136th Symposium of IAU, Los Angeles, U.S.A., Ed M. Morris, Kluwer 
Academic Publishers, Dordrecht, p.157
\bibitem{} Strickland, D. K., Ponman, T. J., Stevens, I. R. 1997, A\&A, 320, 378
\bibitem{} Strong, A. W., Bloemen, J. B. G. M., Dame, T. M., Grenier, I. A., Hermsen, W., 
\bibitem{} Taylor, D., Tyson, Axon, D.J. 1992, MNRAS, 255, 351
\bibitem{} Tomisaka, K., Ikeuchi, S. 1988, ApJ, 330, 695
\bibitem{} Trewhella, M., Davies, J. I., Alton, P. B., Bianchi, S., Madore, B. F. 2000, 
ApJ, 543, 153
\bibitem{} Usero, A., García-Burillo, S., Fuente, A., Martín-Pintado, J.; Rodríguez-Fernández, N. J. 2004 A\&A, 419, 897
\bibitem{} Walter, F.,Weiss, A., Scoville, N. Z. 2002, ApJ, 580, 21
\bibitem{} Wei\ss , A., Neininger, N., H\"{u}ttemeister, S., Klein, U 2001, A\&A, 365, 571
\bibitem{} White, R. L., Becker, R. H. 1992, ApJS, 79, 331
\bibitem{} Zurita, A., Rozas, M., Beckman, J. E. 2000, A\&A, 363, 9
\end{thebibliography}
\end{document}